\documentclass[journal]{IEEEtran}

\usepackage{graphicx}
\usepackage{subcaption}
\usepackage{cite}
\usepackage{amsmath}
\usepackage{bm}
\usepackage{amsfonts}
\usepackage{algorithmic}
\usepackage{array}
\usepackage{url}
\usepackage{stfloats}
\usepackage{enumitem} 
\usepackage{xcolor,soul,framed}
\usepackage{mdwmath}
\usepackage{mdwtab}
\usepackage{lineno,hyperref}

\begin{document}

\title{Joint Semantic-Channel Coding and Modulation for Token Communications}

\author{Jingkai~Ying,~\IEEEmembership{Graduate Student Member,~IEEE,}
        Zhijin~Qin,~\IEEEmembership{Senior Member,~IEEE,}
        Yulong~Feng, \\
        Liejun~Wang,
        and~Xiaoming~Tao,~\IEEEmembership{Senior Member,~IEEE}
\thanks{Parts of this paper have been presented at the IEEE International Conference on Communications, Montreal, Canada, June 2025 \cite{ying2025point}.}
\thanks{Jingkai Ying, Zhijin Qin, and Xiaoming Tao are with the Department of Electronic Engineering, Tsinghua University, Beijing 100084, China, and with the State Key Laboratory of Space Network and Communications, Beijing, 100084, China (e-mail: yjk23@mails.tsinghua.edu.cn; qinzhijin@tsinghua.edu.cn; taoxm@tsinghua.edu.cn).}
\thanks{Yulong Feng is with the State Key Laboratory of Mobile Network and Mobile Multimedia Technology, Shenzhen 518055, China, and with the ZTE Corporation, Shenzhen, 518055, China (e-mail: feng.yulong1@zte.com.cn).}
\thanks{Liejun Wang is with the School of Computer Science and Technology, Xinjiang University, \"{U}r\"{u}mqi 830046, China (e-mail: wljxju@xju.edu.cn).}
}


\maketitle

\begin{abstract}
In recent years, the Transformer architecture has achieved outstanding performance across a wide range of tasks and modalities. Token is the unified input and output representation in Transformer-based models, which has become a fundamental information unit. In this work, we consider the problem of token communication, studying how to transmit tokens efficiently and reliably. Point cloud, a prevailing three-dimensional format which exhibits a more complex spatial structure compared to image or video, is chosen to be the information source. We utilize the set abstraction method to obtain point tokens. Subsequently, to get a more informative and transmission-friendly representation based on tokens, we propose a joint semantic-channel and modulation (JSCCM) scheme for the token encoder, mapping point tokens to standard digital constellation points (modulated tokens). Specifically, the JSCCM consists of two parallel Point Transformer-based encoders and a differential modulator which combines the Gumel-softmax and soft quantization methods. Besides, the rate allocator and channel adapter are developed, facilitating adaptive generation of high-quality modulated tokens conditioned on both semantic information and channel conditions. Extensive simulations demonstrate that the proposed method outperforms both joint semantic-channel coding and traditional separate coding, achieving over 1dB gain in reconstruction and more than 6$\times$ compression ratio in modulated symbols.
\end{abstract}


\section{Introduction}
\label{sec: Introduction}
Semantic communication, which aims to transmit the meaning of information, has recently drawn increasing attention \cite{qin2024ai, qin2024computing}. Leveraging powerful deep learning (DL) models, semantic communication achieves impressive compression ratio and robust performance under extreme channel conditions. However, the absence of unified semantic representations and significant architectural gaps across modalities remain major obstacles to its advancement. Recent revolutionary progress in the field of artificial intelligence (AI) has made it possible to address these challenges. The AI community has witnessed that Transformers are gaining prominence as the models of choice across diverse modalities. After the initial success in natural language processing (NLP) \cite{vaswani2017attention}, Transformers, with their exceptional ability to capture complex relationships, were further applied to other domains such as audio \cite{child2019generating}, vision \cite{osovitskiy2020image}, three-dimensional (3D) data \cite{guo2021pct}, and the multimodal \cite{radford2021learning} community, where they have also achieved remarkable success. Moreover, the parallel computation efficiency and unprecedented scalability of Transformers led to the emergence of large language models (LLMs) and multimodal large language models (MLLMs), profoundly shaping the development paradigm of AI and revolutionizing the way information is processed. These exciting advances provide a new view on semantic communication system design, inspiring us to focus on token transmission based on the Transformers architecture. 

\begin{figure}[!t]
    \centering
    \includegraphics[width=0.40\textwidth]{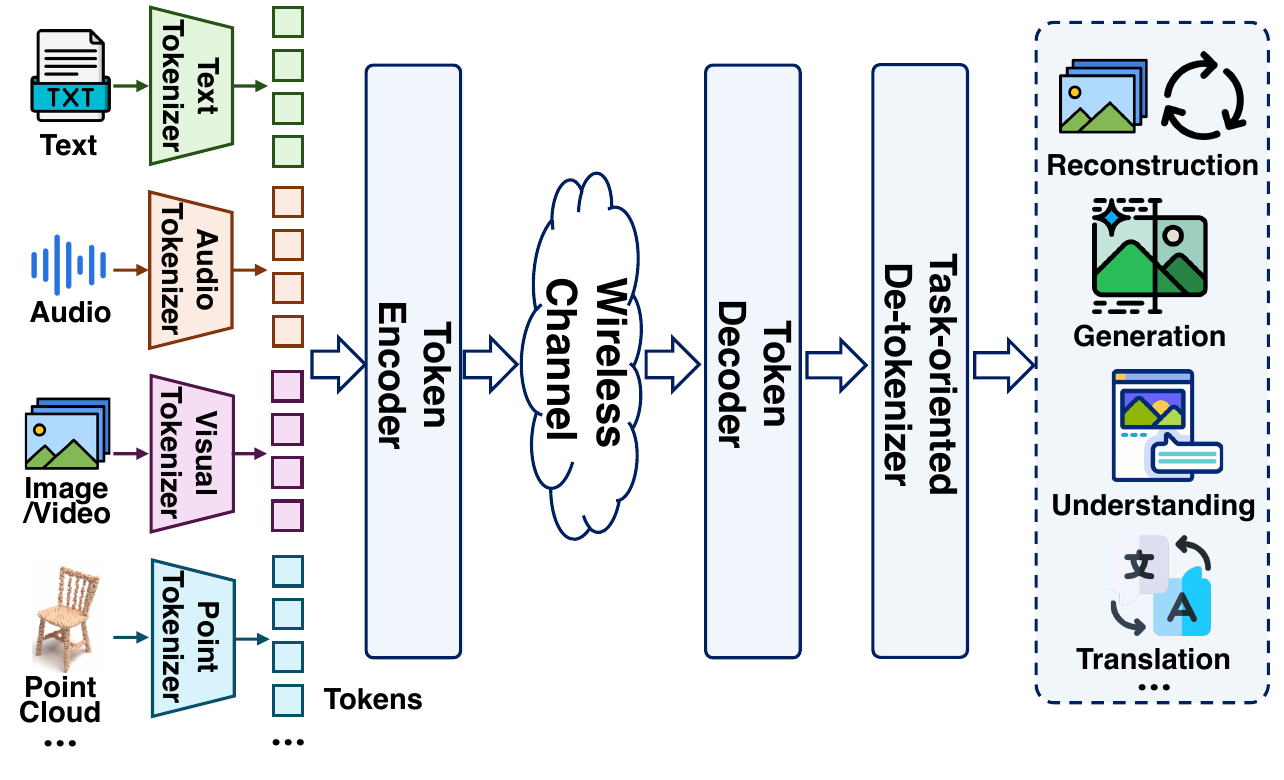}
    \caption{Schematic of token communications \cite{qiao2025token}.}
    \label{fig: token}
\end{figure}

Token, as the unified representation for the input and output of Transformers, is emerging as a new fundamental unit of information. Data from various modalities is converted into tokens for processing \cite{yu2022point, ma2025unitok, hansen2025learnings}. Token/s and Token/J are regarded as key performance metrics for inference speed and energy efficiency, respectively \cite{li2024large}. Moreover, Google introduced the Agent2Agent protocol, further demonstrating the importance of tokens as fundamental units of information \cite{googlea2a}. Currently, extensive efforts are being devoted to token processing to unleash intelligent capacity within the AI community. As important, ensuring efficient and reliable token transmission is also required to establish Integrated AI and Communication \cite{recommendation2023framework}. A conceptual diagram of \textit{token communication} is shown in Fig. \ref{fig: token}. In this framework, source information from different modalities is converted into tokens through tokenizers. The token encoder, based on the Transformers architecture, refines and compresses data based on tokens. The token decoder is used to recover tokens and pass them into the de-tokenizers designed according to tasks at the receiver. 

Recently, there have been some preliminary explorations of token communication, focusing on image transmission \cite{qiao2025token, qiao2025token2}. These pioneering works employ advanced tokenizers to transform images into the token domain and utilize Transformer-based models to predict tokens at the decoder side, which could achieve remarkable improvements in both bandwidth efficiency and semantic fidelity. Furthermore, the utilization of cross-modal information \cite{qiao2025token} and the exploitation of semantic orthogonality \cite{qiao2025token2} reveal new optimization spaces within token communications.

Rather than focusing on image transmission, we apply token communication to point cloud transmission\footnote{Although our model is designed for point cloud, the paradigm of token representing, processing, and transmission can be shared across other modalities. The proposed framework becomes applicable to other modalities when equipped with corresponding tokenizers and Transformer models.}. Point cloud is an important representation of the 3D world, which represents a 3D object through a set of $(x, y, z)$ coordinates, referred to as geometry. Nowadays, point clouds have found increasing applications in fields such as immersive media, autonomous driving, and robotics \cite{valenzise2022immersive}. However, the transmission of point clouds results in a significant increase in data volume, posing substantial challenges to existing communication systems. Dense point cloud sequences can reach a data rate of Gbps level. Therefore, specialized research on point cloud transmission is of paramount importance.

Specifically, we develop a token communication system for point clouds in this paper. To obtain more informative and robust token representations instead of directly using the bitwise form of indices, we design a joint semantic-channel coding and modulation (JSCCM) scheme for the token encoder to generate modulated tokens (constellation points). The joint semantic-channel coding (JSCC) encoder fully exploited the capacity of the Point Transformer \cite{zhang2022transformer}. And the modulator combines the advantage of probabilistic sampling-based methods \cite{bo2024joint}, where the JSCC outputs are interpretable, with the benefit of distance-weighted soft quantization \cite{tung2022deepjscc}, which can explore more available modulation locations. Based on the proposed JSCCM, a rate allocator and a channel adapter are introduced to generate modulated tokens adaptively according to the semantics of point tokens and channel conditions. By integrating the rate allocator and channel adapter into the JSCCM framework, the developed system is no longer fixed-rate and can achieve high-quality transmission under varying channel conditions using a single model.

The main contributions of this paper can be summarized as follows:
\begin{itemize}
\item An end-to-end token communication system for point cloud geometry transmission is developed, which transmits informative and robust modulated tokens over channels. To this end, the JSCCM scheme for the token encoder is devised, which employs two parallel Point Transformers and a differential modulator to map point tokens onto a finite set of digital constellation points.
\item Recognizing the defect of fix-rate transmission, which assigns the same number of constellation points to point clouds with different semantics, the rate allocator is designed. Herein, a masking strategy based on differential cutoff position selection is proposed.
\item To enable the proposed model to adapt to varying channel environments, the channel adapter is introduced. The JSCC outputs representing probabilities of constellation points and channel conditions are concatenated to generate refined JSCC outputs.
\end{itemize}

The remainder of this paper is structured as follows. Section \ref{sec: Related Work} reviews the related work. Section \ref{sec: Framework of the Token Communication System for Point Clouds} introduces the general architecture of the proposed token communication system, identifying the channel models considered and performance metrics considered in this system. The implementation methods of JSCCM, rate allocator, and channel adapter are detailed in Section \ref{sec: Model Design}. Simulation results are provided in Section \ref{sec: Simulation Results}. Section \ref{sec: Conclusion} concludes this paper and provides an outlook for future work.

\textit{Notations}: For a vector $\boldsymbol{x}$, $\| \boldsymbol{x} \|$ denotes its Euclidean norm. For a set $\boldsymbol{X}$, $|\boldsymbol{X}|$ denotes the number of elements in $\boldsymbol{X}$. For a complex number $z$, $|z|$ denotes its norm and $z^{*}$ denotes its complex conjugate. $\mathcal{CN}(\boldsymbol{\mu},\boldsymbol{\Sigma})$ represents the complex Gaussian distribution with mean vector $\boldsymbol{\mu}$ and covariance matrix $\boldsymbol{\Sigma}$. $\mathbb{E}[~\cdot~]$ represents the expectation operation. $\odot$ represents the Hadamard product. $(~\cdot~)^{T}$ and $(~\cdot~)^{H}$ represent transpose and conjugate transpose operations respectively. The spaces of $m \times n$ real and complex matrices are expressed as $\mathbb{R}^{m \times n}$ and $~\mathbb{C}^{m \times n}~$ respectively.

\section{Related Work}
\label{sec: Related Work}

This section reviews the related work on tokenizer design, point cloud semantic communications, and modulation in semantic communications.

\subsection{Tokenizer Design for various modalities}
The concept of tokenizer first emerged in the field of NLP, where it was used to segment text based on spaces, frequency statistics, or other methods. With the widespread adoption of Transformers across various modalities, the concept of the tokenizer has been extended. For instance, dividing an image into patches followed by a linear projection \cite{osovitskiy2020image} is regarded as a form of tokenization. Besides, researchers utilize encoder–decoder models for reconstruction and take outputs of encoders as tokens for downstream tasks \cite{yu2022point, ma2025unitok, hansen2025learnings}. Essentially, a tokenizer aims to produce vector-form tokens that are suitable for further processing by DL models. Depending on whether the outputted tokens are finite or not, tokenizers can be categorized as discrete tokenizers \cite{vaswani2017attention, yu2022point, ma2025unitok} or continuous tokenizers \cite{osovitskiy2020image, liu2023visual}.

\subsection{Point Cloud Semantic Communications}
Currently, motivated by the success of semantic communication in other modalities \cite{xie2021deep, weng2023deep, wang2023wireless}, some researchers have explored DL-enabled semantic communication methods for point cloud transmission systems. In \cite{han2023semantic}, Han \textit{et al.} utilized Point-BERT \cite{yu2022point} as the backbone of the semantic encoder for point cloud classification. Moreover, inspired by the transformer and upsampling-based point cloud compression methods \cite{zhang2022transformer}, Bian \textit{et al.} constructed a Point Transformer \cite{zhao2021point} based semantic communication system \cite{bian2024wireless}. To extract semantic features at different levels for better reconstruction, in \cite{xie2024semantic}, Xie \textit{et al.} employed convolutional networks to obtain image features from projections, while also utilizing PointNet++ \cite{qi2017pointnet++} to extract features from point patches. 

\subsection{Modulation in Semantic Communications}

While the aforementioned researches have preliminarily demonstrated the remarkable performance of semantic communication systems for point cloud transmission, they primarily focused on the design of JSCC structure. It remains challenging to apply these designs to existing digital communication systems. Because current DL-based JSCC utilized in semantic communication systems generates floating-point numbers as outputs. In most settings, every two floating-point numbers are paired to form a constellation point, representing the in-phase and quadrature components, respectively \cite{han2023semantic,bian2024wireless}. This approach is hard to implement with existing hardware and is incompatible with existing communication protocols. 

To tackle this challenge, it is necessary to map the outputs of JSCC to a finite set of channel symbols, enabling semantic communication systems to be compatible with digital communication systems. Methods of quantizing JSCC outputs into bits have been studied in \cite{park2024joint, guo2024digital}, but these methods encounter difficulties in finding an optimal mapping from information source to channel symbols after the introduction of digital modulation due to its non-differentiable nature. Some attempts have been made to find the optimal mapping from JSCC outputs to finite channel symbols in image transmission. In \cite{tung2022deepjscc}, to solve the non-differentiable problem, the constellation points were generated by weighting standard digital constellation points in backpropagation, with the weighting coefficients obtained through a softmax operation on the distances between the JSCC outputs and the standard digital constellation points. And a more intuitive method based on straight-through estimator (STE) was used in \cite{yang2024digital}. Furthermore, by replacing quantization with adding uniform noise in training, a digital semantic communication system for image transmission was developed \cite{zhang2024analog}. In \cite{bo2024joint}, Bo \textit{et al.} adopted a reparameterization method, Gumbel-Softmax \cite{jang2016categorical}, to generate constellation points from a distribution outputted by JSCC. 

\begin{figure*}[!t]
    \centering
    \includegraphics[width=0.75\textwidth]{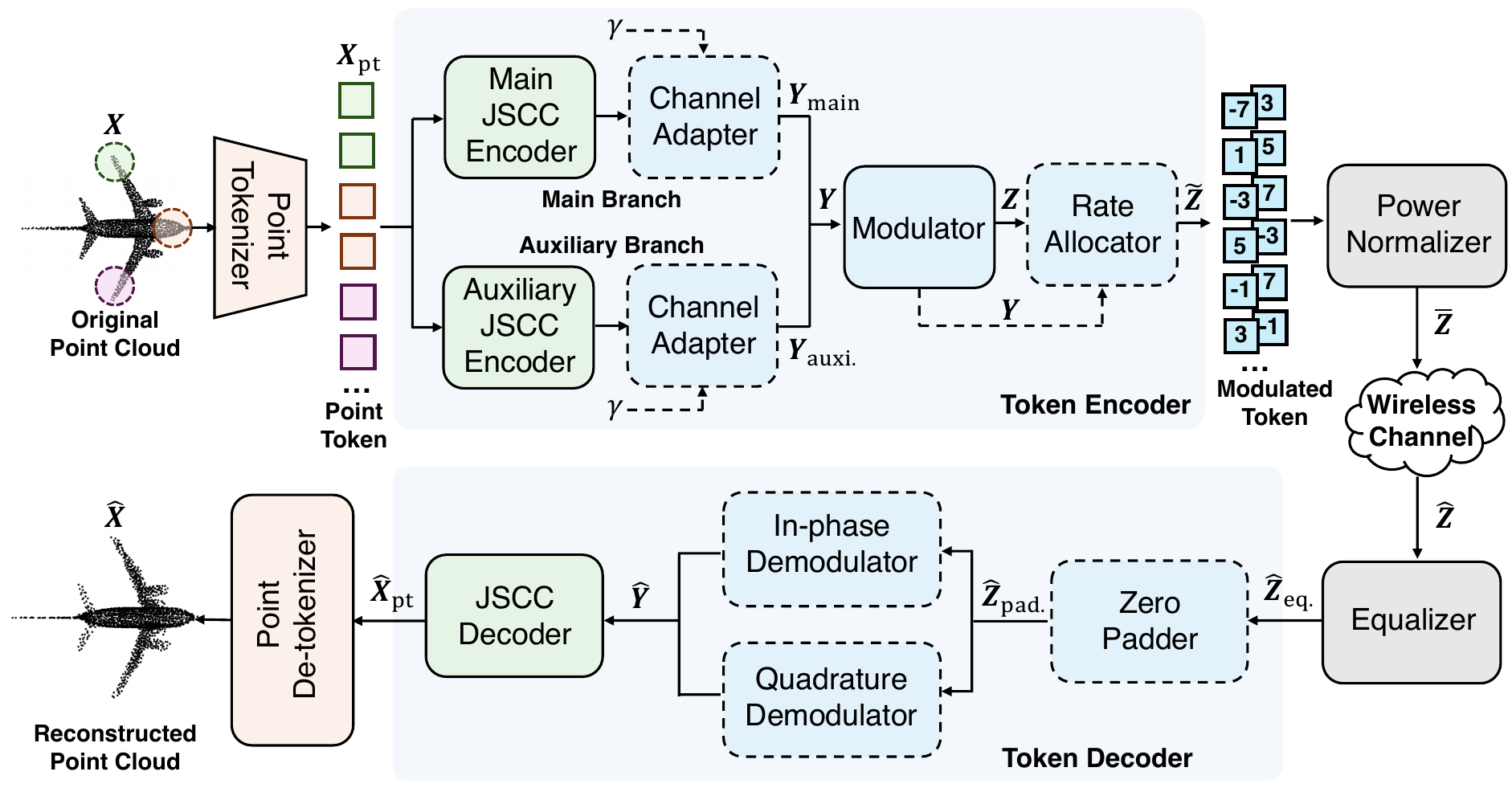}
    \caption{The overall framework of the proposed token communication system for point cloud geometry transmission. Blocks with dashed borders indicate that they are optional. The rate allocator and channel adapter are required when relevant adaptive characteristics are needed. If Rayleigh fading channels are considered, the equalizer will be used.}
    \label{fig: framework}
\end{figure*}

\section{Framework of the Token Communication System for Point Clouds}
\label{sec: Framework of the Token Communication System for Point Clouds}

In this section, we provide an overview of the proposed token communication system for point clouds. First, the inputs, outputs, functions of each system model, and their connection relationship are introduced. Subsequently, the evaluation metrics are presented.

\subsection{Formulation of System Models}
The overall framework of the proposed point cloud geometry transmission system is illustrated in Fig. \ref{fig: framework}. At the transmitter, a point tokenizer converts point patches into point tokens represented by vectors, and a token encoder further refines point tokens and transforms tokens into a modulated form for transmission. The token encoder comprises two parallel JSCC encoders for producing the probabilities of constellation point positions, which further guide the modulator to generate modulated tokens (constellation points). This parallel design naturally forms a main branch and an auxiliary branch, enabling the generation of multi-level point cloud features. Additionally, the token encoder can embed rate allocator and channel adapter to achieve corresponding adaptive capabilities. At the receiver, the received constellation points are demodulated on the in-phase and quadrature paths separately. The semantic features obtained from demodulation are then fed into the JSCC decoder and subsequently passed through the de-tokenizer for the final reconstruction task. When Rayleigh fading and rate adaptive transmission are considered, the corresponding equalization and constellation points padding procedure will be conducted at the receiver. 

To be specific, since only the geometry information is considered, a point cloud can be expressed as $\boldsymbol{X} =\left \{ \boldsymbol{x}_{i}\right \} $, $i=1,\cdots,N$, where \( N \) is the number of points in the point cloud, and $\boldsymbol{x}_{i} \in \mathbb{R}^{3} $ represents each point's three-dimensional coordinates. The raw point cloud is first processed by the point tokenizer, which reduces the number of points in the point cloud and enhances feature dimensions, resulting in $\boldsymbol{X}_\text{pt}=\left \{ \boldsymbol{x{}'}_{i}\right \} $, $i=1,\cdots,N{}' $, where $N{}'$ denotes the number of point tokens and $\boldsymbol{x}{}'_{i} \in \mathbb{R}^{C{}'} $ is the embedding of the token. $C{}'$ is the feature dimension of the point token. Subsequently, $\boldsymbol{X}_\text{pt}$ is fed into two parallel DL-based JSCC encoders, namely the main JSCC encoder and the auxiliary JSCC encoder, to obtain more informative point cloud semantic features. If channel adaptation is required, two independent channel adapters are incorporated into the main branch and auxiliary branch to adjust the outputs of the JSCC encoders based on the channel condition $\boldsymbol{\gamma}$. The output of the main branch $\boldsymbol{Y}_\text{main}$  can be obtained by
\begin{align}
\boldsymbol{Y}_\text{main}= \mathcal{C}_\text{main}\left ( \mathcal{S}_\text{main} \left ( \boldsymbol{X}_\text{pt} ;\boldsymbol{\alpha} _\text{main}\right ); \boldsymbol{\beta} _\text{main} , \boldsymbol{\gamma}\right ), 
\end{align}
where $\mathcal{S}_\text{main}\left (~\cdot~; \boldsymbol{\alpha}_\text{main} \right)$, $\mathcal{C}_\text{main}\left (~\cdot~; \boldsymbol{\beta}_\text{main}, \boldsymbol{\gamma} \right)$ denotes the main JSCC encoder and channel adapter with trainable parameters $\boldsymbol{\alpha}_\text{main}$, $\boldsymbol{\beta}_\text{main}$ correspondingly. $\boldsymbol{\gamma}$ is the channel conditions. In the same way, for the auxiliary branch,  $\boldsymbol{Y}_\text{auxi.}=\mathcal{C}_\text{auxi.}\left ( \mathcal{S}_\text{auxi.} \left ( \boldsymbol{X}_\text{pt} ;\boldsymbol{\alpha} _\text{auxi.}\right ); \boldsymbol{\beta} _\text{auxi.} , \boldsymbol{\gamma}\right )$. Then $\boldsymbol{Y}_\text{main}$, $\boldsymbol{Y}_\text{auxi.}$ are concatenated as $\boldsymbol{Y}$ and sent to the modulator to generate modulated tokens $\boldsymbol{Z} \in \mathbb{C}^{N_\text{mod.}} $. If the number of transmitted constellation points needs to be dynamically adjusted based on the semantic features of point clouds, then before transmission, the rate allocator will generate a mask based on JSCC outputs to discard a part of the modulated tokens. After the rate allocation operation, the resulting constellation points are represented as $\boldsymbol{\tilde{Z} } \in \mathbb{C}^{N_\text{send}} \left( N_\text{send} \le N_\text{mod.} \right )$. The above process can be formulated as
\begin{align}
\boldsymbol{\tilde{Z} } = \mathcal{R} \left ( \mathcal{M} \left ( \boldsymbol{Y}; \boldsymbol{\xi} \right ); \boldsymbol{Y}, \boldsymbol{\zeta} \right ),
\end{align}
where $\boldsymbol{Y}=\operatorname{concat}\left ( \boldsymbol{Y}_\text{main}, \boldsymbol{Y}_\text{auxi.}\right )$. $\mathcal{M} \left (~\cdot~; \boldsymbol{\xi} \right)$ and $\mathcal{R} \left (~\cdot~; \boldsymbol{Y},\boldsymbol{\zeta} \right)$ are modulator and rate allocator with parameters $\boldsymbol{\xi}$ and $\boldsymbol{\zeta}$. It should be noted that $\boldsymbol{\zeta}$ is trainable, while $\boldsymbol{\xi}$ represents the available standard digital constellation points that are manually set in advance as a token codebook. The output of the power normalizer is denoted as $\bar{\boldsymbol{Z}} \in \mathbb{C}^{N_\text{send}}$ with targeted power. $\bar{\boldsymbol{Z}}$ will be transmitted over the wireless channel. 

We consider both the additive white Gaussian noise (AWGN) channel and the Rayleigh fading channel. At the receiver, the received constellation points are denoted as $\hat{\boldsymbol{Z}}$. As for the AWGN scenario, $\hat{\boldsymbol{Z}}$ can be represented as 
\begin{align}
\hat{\boldsymbol{Z}} = \bar{\boldsymbol{Z}} + \boldsymbol{n},
\end{align}
where $\boldsymbol{n}$ is the Gaussian noise with noise power $P_\text{noise}$, and $\boldsymbol{n}\sim \mathcal{CN}(0,P_\text{noise}\mathbf{I})$. $\mathbf{I}$ is a $N_\text{send} \times N_\text{send}$ identity matrix. The received signal power can be calculated as 
\begin{align}
P_\text{signal}=\frac{\mathbb{E} \left[ \| \boldsymbol{\bar{Z} } \|_2^2 \right]  }{N_\text{send}}. 
\end{align}
Correspondingly, the signal-to-noise ratio (SNR) in decibels for the AWGN channel can be calculated as
\begin{align}
\label{snr}
\text{SNR} = 10\log_{10}{\frac{P_\text{signal} }{P_\text{noise} } } . 
\end{align}
As for the Rayleigh fading scenario, $\hat{\boldsymbol{Z}}$ can be obtained as 
\begin{align}
\hat{\boldsymbol{Z}} = h\cdot\bar{\boldsymbol{Z}} + \boldsymbol{n},
\end{align}
where $h$ is the channel gain between the transmitter and the receiver. The definition of $\boldsymbol{n}$ is the same as in the AWGN scenario. The received signal power should be calculated as 
\begin{align}
P_\text{signal}=\frac{\mathbb{E} \left[ \| h\cdot\boldsymbol{\bar{Z} } \|_2^2 \right]  }{N_\text{send}}. 
\end{align}
The SNR in decibels for the Rayleigh fading case is also calculated following (\ref{snr}). For both types of channels, SNRs are fed into the channel adaptation module as the channel condition $\boldsymbol{\gamma}$.

The receiver will perform zero forcing (ZF) equalization in the Rayleigh fading scenario as
\begin{align}
\hat{\boldsymbol{Z}}_{\text{eq.}}=\frac{h^{*}}{|h|^{2}} \cdot \hat{\boldsymbol{Z}},
\end{align}
where $\hat{\boldsymbol{Z}}_{\text{eq.}}$ is the output of the equalizer. And when the rate allocator is included at the transmitter, the receiver will conduct zero padding. By padding zeros to restore the symbol number to ${N_\text{mod.}}$, the dimensions of $\hat{\boldsymbol{Z}}_{\text{pad.}}$ can match the subsequent demodulation models.  Subsequently, $\hat{\boldsymbol{Z}}_{\text{pad.}}$ is divided into the in-phase and quadrature paths for demodulation individually, completing the transformation from the constellation points $\hat{\boldsymbol{Z}}_{\text{pad.}}$  to the semantic features $\hat{\boldsymbol{Y}}$.  $\hat{\boldsymbol{Y}}$ is further processed in the JSCC decoder to get a downsampled point cloud $\hat{\boldsymbol{X}}_\text{pt}$. The above process can be represented as
\begin{align}
\hat{\boldsymbol{X}}_\text{pt} =\mathcal{S}^{-1}\left ( \mathcal{M}^{-1} \left ( \mathcal{R}^{-1}\left ( \hat{\boldsymbol{Z}} \right );\boldsymbol{\kappa}    \right );\boldsymbol{\eta}   \right ),
\end{align}
where $\mathcal{R} ^{-1}\left (~\cdot~ \right )$ represents the padding process, which does not require a mask from the transmitter. $\mathcal{M} ^{-1}\left (~\cdot~ ; \boldsymbol{\kappa} \right )$ represents demodulator with trainable parameters $\boldsymbol{\kappa}$ and  $\mathcal{S} ^{-1}\left (~\cdot~ ; \boldsymbol{\eta} \right )$  represents JSCC decoder with trainable parameters $\boldsymbol{\eta}$. $\hat{\boldsymbol{X}}_\text{pt}$ is upsampled in the point de-tokenizer to get the reconstructed point cloud $\hat{\boldsymbol{X}}$ ultimately.

\subsection{Performance Metrics}
Two commonly used metrics for measuring the geometric fidelity of point clouds, namely $\text{D1}$ and $\text{D2}$ \cite{ISO2020PointCloud}, are adopted to evaluate the end-to-end performance of the proposed communication system. To be specific, the $\text{D1}$ metric based the original point cloud $\boldsymbol{X}$ and the reconstructed point cloud $\hat{\boldsymbol{X}}$ are defined as
\begin{align}
\text{D1} = \max \left\{ e_{1,\boldsymbol{X} \to \hat{\boldsymbol{X}}} , e_{1,\hat{\boldsymbol{X}} \to \boldsymbol{X}} \right \},
\end{align}
where $e_{1,\boldsymbol{X} \to \hat{\boldsymbol{X}}}$ and $e_{1,\hat{\boldsymbol{X}} \to \boldsymbol{X}}$ can be calculated as (\ref{D1}).  $e_{1,\boldsymbol{X} \to \hat{\boldsymbol{X}}}$ and $e_{1,\hat{\boldsymbol{X}} \to \boldsymbol{X}}$ represent the average point-to-point distance square between two point clouds. This point-to-point distance is defined as the minimum Euclidean distance from a specific point in one point cloud to points in the other point cloud.
\begin{align}
\label{D1}
\notag
&e_{1,\boldsymbol{X} \to \hat{\boldsymbol{X}}} = \frac{1}{|\boldsymbol{X}|} \sum_{\boldsymbol{x} \in \boldsymbol{X}} \min_{\hat{\boldsymbol{x}} \in \hat{\boldsymbol{X}}} \| \boldsymbol{x} - \hat{\boldsymbol{x}} \|_2^2 ,
\\&e_{1,\hat{\boldsymbol{X}} \to \boldsymbol{X}} = \frac{1}{|\hat{\boldsymbol{X}}|} \sum_{\hat{\boldsymbol{x}} \in \hat{\boldsymbol{X}}} \min_{\boldsymbol{x} \in \boldsymbol{X}} \| \hat{\boldsymbol{x}} - \boldsymbol{x} \|_2^2.
\end{align}
The peak signal-to-noise ratio (PSNR) version in decibels is utilized more frequently. \text{D1 PSNR} can be calculated as
\begin{align}
\label{D1_PSNR}
\text{D1 PSNR} = 10\log_{10}{\frac{3P^{2}}{\text{D1}} }.
\end{align}
where $P$ is the peak value of the point cloud coordinates. 

$\text{D2}$ is defined similarly as
\begin{align}
\text{D2} = \max \left\{ e_{2,\boldsymbol{X} \to \hat{\boldsymbol{X}}} , e_{2,\hat{\boldsymbol{X}} \to \boldsymbol{X}} \right \},
\end{align}
where $e_{2,\boldsymbol{X} \to \hat{\boldsymbol{X}}}$ and $e_{2,\hat{\boldsymbol{X}} \to \boldsymbol{X}}$ are calculated based on the average point-to-plane distance. This point-to-plane distance is the shortest Euclidean distance from a specific point in one point cloud to the planes where points in the other point cloud lie. Taking $e_{2,\boldsymbol{X} \to \hat{\boldsymbol{X}}}$ as an example, it can be calculated as 
\begin{align}
\label{2ex2x_hat}
e_{2,\boldsymbol{X} \to \hat{\boldsymbol{X}}} = \frac{1}{|\boldsymbol{X}|} \sum_{\boldsymbol{x} \in \boldsymbol{X}} \cos^2 \theta_{\boldsymbol{x} \hat{\boldsymbol{x}}} \min_{\hat{\boldsymbol{x}} \in \hat{\boldsymbol{X}}} \| \boldsymbol{x} - \hat{\boldsymbol{x}} \|_2^2 ,
\end{align}
in which $\theta_{\boldsymbol{x} \hat{\boldsymbol{x}}}$ is the angle between vector $\boldsymbol{x} - \hat{\boldsymbol{x}}$ and the normal vector of $\hat{\boldsymbol{x}}$. $e_{2,\hat{\boldsymbol{X}} \to \boldsymbol{X}}$ can be calculated in the same way by swapping the position of $\boldsymbol{X}$ and $\hat{\boldsymbol{X}}$ and $\boldsymbol{x}$ and $\hat{\boldsymbol{x}}$. And the PSNR value of D2 is defined by replacing D1 with D2 in (\ref{D1_PSNR}). It should be highlighted that the calculation of D1 is simpler because it does not include normal vectors, while D2 more accurately reflects the perceptual characteristics of the human visual system \cite{valenzise2022immersive}.

\section{Model Design}
\label{sec: Model Design}

This section presents the point tokenizer first. Then it focuses on the proposed JSCCM scheme for the token encoder, including two parallel JSCC encoders and the differentiable modulator. Besides, two adaptive modules, namely the rate allocator and channel adapter, are introduced. Finally, the demodulator, JSCC decoder, and point de-tokenizer are presented.

\subsection{Point Tokenizer}

We adopt Set Abstraction, first introduced in PointNet++ \cite{qi2017pointnet++}, to perform point tokenization. Set Abstraction effectively aggregates the features in a point patch, resulting in subsampled points with higher feature dimensions. Each of these subsampled points is treated as a point token, with its features serving as the embedding of the point token. In this way, each point token represents the structural features of a local region in the point cloud. When processed by the subsequent Point Transformer \cite{zhao2021point}, the geometric relationships among the point tokens are further captured, resulting in tokens with enhanced representation capabilities.

Set Abstraction consists of three key layers: \textit{Sampling Layer}, \textit{Grouping Layer}, and \textit{PointNet Layer}. In the \textit{Sampling Layer}, the farthest point sampling (FPS) algorithm is applied. It starts by randomly selecting a point $\boldsymbol{x}_{c_{1}}$ from the original point cloud 
$\boldsymbol{X} =\left \{ \boldsymbol{x}_{1}, \dots, \boldsymbol{x}_{N}\right \} $ as a centroid. Then, the point $\boldsymbol{x}_{c_{2}}$ that is farthest from $\boldsymbol{X}_\text{centroid} = \left \{\boldsymbol{x}_{c_{1}}\right \}$ is selected from the remaining points $\boldsymbol{X} \setminus \left \{ \boldsymbol{x}_{c_{1}}\right \}$ and added to the centroid set $\boldsymbol{X}_\text{centroid}$. This process is iterated, where each time, the point $\boldsymbol{x}_{c_{j}}$ farthest from the centroid set $\boldsymbol{X}_\text{centroid} = \left \{ \boldsymbol{x}_{c_{1}},\dots,\boldsymbol{x}_{c_{j-1}}\right \}$ is selected from the remaining points $\boldsymbol{X} \setminus \left \{ \boldsymbol{x}_{c_{1}},\dots,\boldsymbol{x}_{c_{j-1}}\right \}$. The ultimate output of the \textit{Sampling Layer} is $\boldsymbol{X}_\text{centroid} = \left \{ \boldsymbol{x}_{c_{1}},\dots,\boldsymbol{x}_{c_{N{}'}}\right \}$. It is worth noting that the distance from a point to a point set is defined as the minimum Euclidean distance between this point and all points in the set. In the \textit{Grouping Layer}, the ball query algorithm is conducted to find the $K$ neighboring points \footnote{In implementation, if there are fewer than $K$ points within the radius, the point corresponding to the smallest index within the radius is used for padding. If there are more than $K$ points, the points are sorted in descending order of their indices, and the top $K$ points are selected.} of each centroid $\boldsymbol{x}_{c_{j}}$ within a specified radius. In the local region corresponding to each centroid point $\boldsymbol{x}_{c_{j}}$, the coordinates and features of the $K$ neighboring points are grouped. Since only point geometry is considered, features in the Set Abstraction for point tokenizer are simply coordinates. In the \textit{PointNet Layer}, the coordinates of the neighboring points are transformed into a local frame relative to the corresponding centroid points. Subsequently, as in PointNet \cite{Qi_2017_CVPR}, a $1\times1$ convolution with shared parameters across different local regions is utilized to adjust the dimension of grouped features. Finally, by performing max pooling within each local region, the output of Set Abstraction is obtained. 

\begin{align}
\label{flow}
\notag
&\underbrace{(N,3) \xrightarrow{\text{FPS}} (N{}' ,3)}_{\text{Sampling Layer}}\underbrace{\xrightarrow{\text{ball query}}(N{}' ,K,C) \xrightarrow{\text{group}}(N{}' ,K,C+3)}_{\text{Grouping Layer}} \\ 
&\underbrace{\xrightarrow{\text{diff.}} (N{}' ,K,C+3)\xrightarrow{1\times1\text{ conv.}}(N{}' ,K,C{}')\xrightarrow{\text{pool.}}(N{}' ,C{}')}_{\text{PointNet Layer}}.
\end{align}

The dimension changes of the input and output can be represented by (\ref{flow}). In (\ref{flow}), numbers in parentheses represent dimensions of inputs and outputs. $C$ is the feature dimension of points. Since we only consider the transmission of point cloud geometry, the value of $C$ for the point tokenizer is set to 3. $C{}'$ is the feature dimension of points adjusted by PointNet.

\begin{figure*}[!t]
    \centering
    \includegraphics[width=0.75\textwidth]{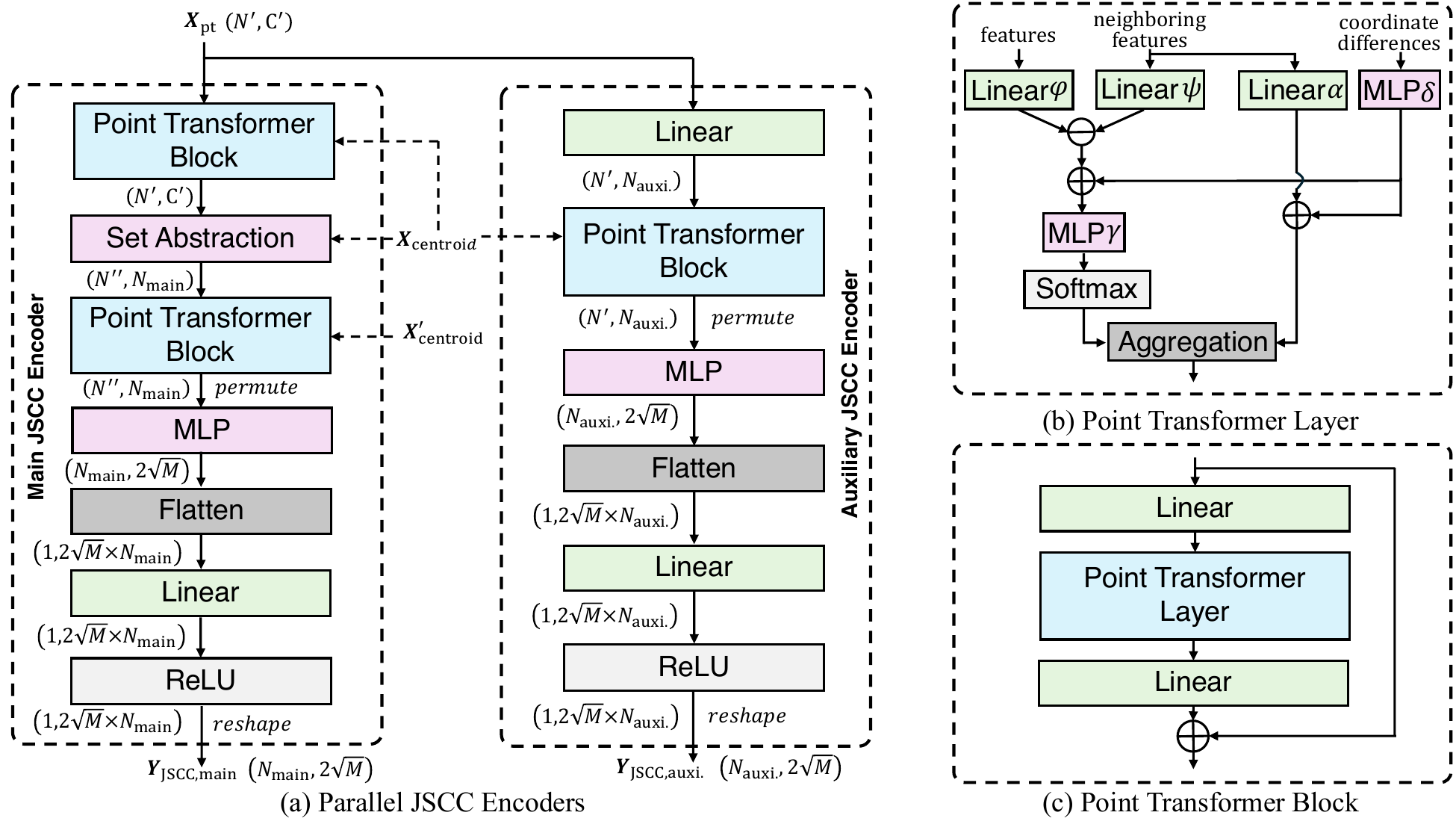}
    \caption{The model architecture of JSCC encoders. The numbers in brackets are dimension information. (a) The diagram of two parallel JSCC encoders, including a main JSCC encoder and an auxiliary encoder. (b) The diagram of a Point Transformer layer, which conducts vector attention. (c) The diagram of a Point Transformer Block.}
    \label{fig: encoder}
\end{figure*}

\subsection{JSCCM Scheme for the Token Encoder} 
\label{JSCCM}
The JSCCM scheme is designed for the token encoder to generate more informative and robust tokens. JSCCM consists of two parallel JSCC encoders implemented with Point Transformer \cite{zhao2021point} and a differentiable modulation method that combines Gumbel-Softmax \cite{bo2024joint} and soft quantization \cite{tung2022deepjscc}.

As for JSCC encoders, the detailed structure of the two parallel JSCC encoders and the illustration of Point Transformer are shown in Fig. \ref{fig: encoder}. The encoders have two branches: the main JSCC encoder  $\mathcal{S}_\text{main}\left (~\cdot~; \boldsymbol{\alpha}_\text{main} \right)$ and the auxiliary JSCC encoder $\mathcal{S}_\text{auxi.}\left (~\cdot~; \boldsymbol{\alpha}_\text{auxi.} \right)$, which are used to extract different levels of geometric features from the point cloud. Both JSCC encoders take the point tokens $\boldsymbol{x}{}'_{i} \in \boldsymbol{X}_\text{pt} $ obtained from Set Abstraction as input, and further characterize the relationships between point tokens through the Point Transformer to enhance the representation capability of the tokens. 

The Point Transformer no longer uses dot products to compute attention scores. Instead, it computes the difference between query and key and then applies a multilayer perception (MLP) for a nonlinear transformation to obtain the attention scores, which is a vector attention mechanism. Additionally, since point clouds inherently contain positional information, the position encoding of the Point Transformer is obtained by computing the difference of the point cloud coordinates, followed by a nonlinear transformation through an MLP. The first Point Transformer layer in the main branch can be formulated as
\begin{align}
\notag
\boldsymbol{x{}''}_{i}=\sum_{\boldsymbol{x{}'}_{j} \in \tilde{\boldsymbol{X}}_\text{j,pt}} &\text{softmax}\left(\frac{\gamma\left(\varphi\left(\boldsymbol{x{}'}_{i}\right)-\psi\left(\boldsymbol{x{}'}_{j}\right)+\delta(\boldsymbol{x{}}_{i}-\boldsymbol{x{}}_{j})\right)}{\sqrt{C{}'}}\right) \\&\odot\left(\alpha\left(\boldsymbol{x{}'}_{j}\right)+\delta(\boldsymbol{x{}}_{i}-\boldsymbol{x{}}_{j})\right),
\end{align}
where $\tilde{\boldsymbol{X}}_\text{j,pt} \subseteq \boldsymbol{X}_\text{pt}$, and $\tilde{\boldsymbol{X}}_\text{j,pt}$ is obtained by indexing $\boldsymbol{X}_\text{pt}$. The indices are acquired by conducting k-nearest neighbors (kNN) on $\boldsymbol{x}_{c_{j}}$ over $\boldsymbol{X}_\text{centroid} = \left \{ \boldsymbol{x}_{c_{1}},\dots,\boldsymbol{x}_{c_{N{}'}}\right \}$. $\varphi$, $\psi$, and $\alpha$ represent linear layers projecting point tokens to get query, key, and value, respectively. Here $\boldsymbol{x{}}_{i}, \boldsymbol{x{}}_{j} \in \boldsymbol{X}$ are the point coordinates for $\boldsymbol{x{}}{}'_{i}$ and $\boldsymbol{x{}}{}'_{j}$. $\gamma$ and $\delta$ are both MLPs, which consist of two linear layers and one linear rectification function (ReLU). To control the feature dimensions and alleviate the vanishing gradient problem, a linear layer is added before and after the Point Transformer layer, and the input and output are linked through a residual connection to form a complete Point Transformer block.

As illustrated in Fig. \ref{fig: encoder}, compared to the auxiliary JSCC encoder, the main JSCC encoder includes an additional Set Abstraction operation to further aggregate features and an additional Point Transformer to further capture relationships across different local regions. In addition to the core Point Transformer, both parallel JSCC encoders incorporate MLPs to adjust the feature dimensions, enabling the outputs to be compatible with the subsequent modulation module. The adjacent flatten and linear layers in each encoder are designed to distribute the semantic information of the point cloud across different constellation points as much as possible. The final outputs of the two parallel JSCC encoders are formatted as logits representing the probability of each constellation point location. They can be represented as $\boldsymbol{Y}_\text{JSCC, main}=\mathcal{S}_\text{main}\left (\boldsymbol{X}_\text{pt}; \boldsymbol{\alpha}_\text{main} \right) \in \mathbb{R}^{N_{\text{main}} \times 2\sqrt{M}}$, $\boldsymbol{Y}_\text{JSCC, auxi.}=\mathcal{S}_\text{auxi.}\left (\boldsymbol{X}_\text{pt}; \boldsymbol{\alpha}_\text{auxi.} \right) \in \mathbb{R}^{N_{\text{auxi.}} \times 2\sqrt{M}}$. $N_{\text{main}}$, $N_{\text{auxi.}}$ denote the number of modulated tokens in each branch. $M$ refers to the number of constellation points in quadrature amplitude modulation (QAM), which decides the available choices in the token codebook. Similar to $\boldsymbol{X}_\text{centroid}$, $\boldsymbol{X{}'}_\text{centroid}$ represents $N{}''$ centroid points, obtained by sampling of Set Abstraction in the main JSCC encoder.

\begin{figure}[!t]
    \centering
    \includegraphics[width=0.40\textwidth]{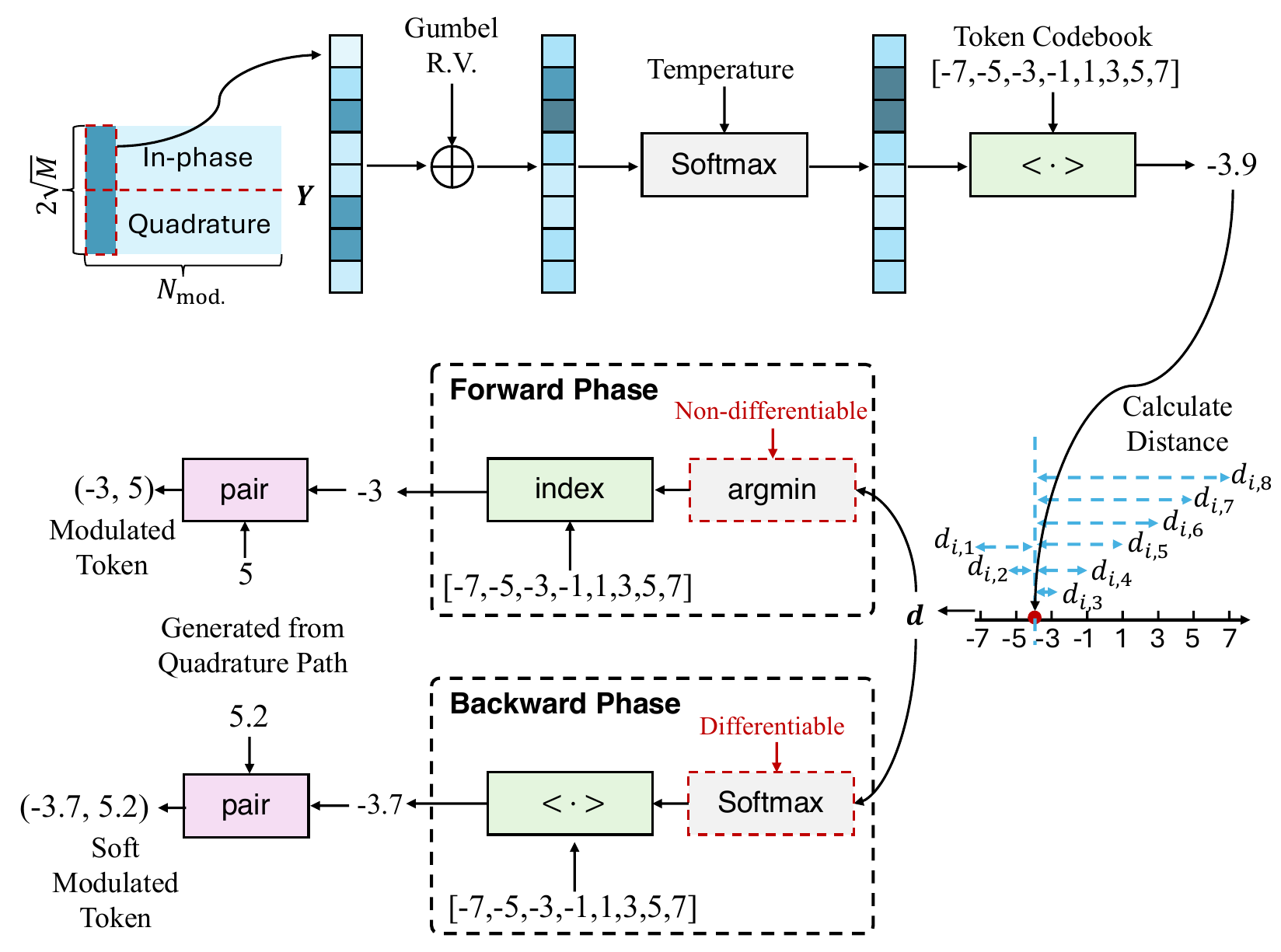}
    \caption{An illustrative example of the proposed differentiable modulation method. 64-QAM is used in this example, which means $M=64$.}
    \label{fig: modulator}
\end{figure}

As for the differentiable modulation method, the process of modulation is demonstrated in Fig. \ref{fig: modulator}. Since the rate allocator and channel adapter in the proposed communication system are optional, we skip the channel adapter for now to present the concepts smoothly.$M$-QAM, which means there are $M$ constellation points arranged in a square, is considered for the modulation scheme. The input of the modulator is $\boldsymbol{Y} \in \mathbb{R}^{N_{\text{mod.}} \times 2\sqrt{M}}$ where $N_{\text{mod.}}$ denotes the total number of modulated tokens generated before rate allocation. Instead of converting bits into constellation points, the modulator here generates constellation points based on the semantic features $\boldsymbol{Y} $, which represent the probabilities of constellation point locations. After passing through the JSCC encoders and the modulator, the point tokens to be transmitted are converted into modulated tokens with fewer symbols, rich semantics, and suitability for transmission.

We denote $\boldsymbol{y}_{i} \in \mathbb{R}^{1 \times 2\sqrt{M}}$ as the $i$-th row of $\boldsymbol{Y}$, which contains the positional information of the $i$-th constellation point. Furthermore, let $\boldsymbol{y}_{i}=(\boldsymbol{y}_{\text{I},i},\boldsymbol{y}_{\text{Q},i})$ which means the first $\sqrt{M}$ components of $\boldsymbol{y}_{i}$ represent logits for the in-phase position of the constellation point, while the last $\sqrt{M}$ components represent logits for the quadrature position. 

Taking the generation of in-phase position of the $i$-th constellation point as an example, in the forward phase, we use the Gumbel-Softmax method to obtain a soft output of the constellation point position probabilities $\boldsymbol{t}_{\text{I},i}$, as
\begin{align}
\boldsymbol{t}_{\text{I},i}(j)=\frac{\exp \left(\left[\tau_{j}+\boldsymbol{y}_{\text{I},i}(j)\right] / T\right)}{\sum_{k=1}^{\sqrt{M}} \exp \left(\left[\tau_{k}+\boldsymbol{y}_{\text{I},i}(k)\right] / T\right)},
\end{align}
where $\boldsymbol{t}_{\text{I},i}(j)$ is the $j$-th entry of $\boldsymbol{t}_{\text{I},i}$ and $\boldsymbol{y}_{\text{I},i}(j)$ is the $j$-th element in $\boldsymbol{y}_{\text{I},i}$. $\tau_{j}$ is sampled from $\operatorname{Gumbel}(0,1)$. This is the Gumbel-Softmax trick \cite{jang2016categorical}, which allows a discrete distribution to be reparameterized for sampling. It transforms the process of sampling from a discrete distribution into the process of finding the maximum element ($\operatorname{argmax}$) of a vector with added randomness. And the $\operatorname{softmax}$ operation is used to approximate the non-differential $\operatorname{argmax}$ operation. $T$ is the temperature hyperparameter, which is utilized to control the steepness of the distribution of $\boldsymbol{t}_{\text{I},i}$. 

Then we define 
$\boldsymbol{c}=(c_{1},c_{2},\ldots,c_{\sqrt{M}})$ as the token codebook, consisting of the standard coordinates of an $M$-QAM modulation scheme. Based on $\boldsymbol{t}_{\text{I},i}$ and $\boldsymbol{c}$, the inner product can be calculated to generate an initial position by
\begin{align}
z_{\text{I},i} = < \boldsymbol{t}_{\text{I},i},\boldsymbol{c}>= \boldsymbol{t}_{\text{I},i} \cdot \boldsymbol{c}^{\text{T}}.
\end{align}

Distances between $z_{\text{I},i}$ and $\boldsymbol{c}$ can be further calculated as$\boldsymbol{d}_{i}=(d_{i,1},d_{i,2},\ldots,d_{i,\sqrt{M}})=(|c_{1}-z_{\text{I},i}|,|c_{2}-z_{\text{I},i}|,\ldots,|c_{\sqrt{M}}-z_{\text{I},i}|)$. And based on the distances between the initial position and standard coordinates of $M$-QAM, the quantization is performed as
\begin{align}
\label{product}
\bar{z}_{\text{I},i}
\notag
&=<\operatorname{one-hot}\left(\underset{j \in\{1, \ldots, \sqrt{M}\}}{\operatorname{argmin}}d_{i,j}\right), \boldsymbol{c}>,\\& = \operatorname{one-hot}\left(\underset{j \in\{1, \ldots, \sqrt{M}\}} {\operatorname{argmin}}d_{i,j}\right) \cdot \boldsymbol{c}^{T},
\end{align}
where $\bar{z}_{\text{I},i}$ is the final output corresponding to the in-phase coordinate of the $i$-th constellation point. Similarly, the coordinate of the $i$-th constellation point $\boldsymbol{z}_{\text{Q},i}$ is generated, followed by the aforementioned process with $\boldsymbol{y}_{\text{Q},i}$ as inputs. $z_{\text{I},i}$ and $z_{\text{Q},i}$ will be paired to form a constellation points. Thus far, the modulator maps the logits $\boldsymbol{y}_{i}$ to the modulated token $\boldsymbol{z}_{i}=(z_{\text{I},i},z_{\text{Q},i})$.

Since the $\operatorname{argmin}$ operation in (\ref{product}) is non-differential, in the backward phase, a soft quantization method is used and the soft output corresponding to the in-phase coordinate of the $i$-th constellation point $\tilde{z}_{\text{I},i}$ is calculated by
\begin{align}
\label{soft}
\tilde{z}_{\text{I},i}=\sum_{k=1}^{\sqrt{M}} \frac{\exp \left(-d_{i,k} / T\right)}{\sum_{j=1}^{\sqrt{M}} \exp \left(-d_{i,j} / T\right)}\cdot c_{k}.
\end{align}
In implementation, the computations of the forward and backward stages can be integrated as follows
\begin{align}
z^{\text{I},i}_{\text{output}} =\operatorname{detach}(\bar{z}_{\text{I},i}-\tilde{z}_{\text{I},i}) + \tilde{z}_{\text{I},i},
\end{align}
where $z^{\text{I},i}_{\text{output}}$ can represent the output of the $i$-th constellation points in In-phase for both forward phase and backward phase. Because the $\operatorname{detach}$ operation has no effect during the forward phase, so $z^{\text{I},i}_{\text{output}} = \bar{z}_{\text{I},i}$. While during the backward phase, the $\operatorname{detach}$ will detach $(\bar{z}_{\text{I},i}-\tilde{z}_{\text{I},i})$ from the computation graph. The gradient will propagate through the differentiable $z^{\text{I},i}_{\text{output}} = \tilde{z}_{\text{I},i}$. This logic can be easily implemented using PyTorch.

\subsection{Rate Allocator and Channel Adapter} 

\begin{figure}[!t]
    \centering
    \begin{subfigure}{0.40\textwidth}
        \centering
        \includegraphics[width=\textwidth]{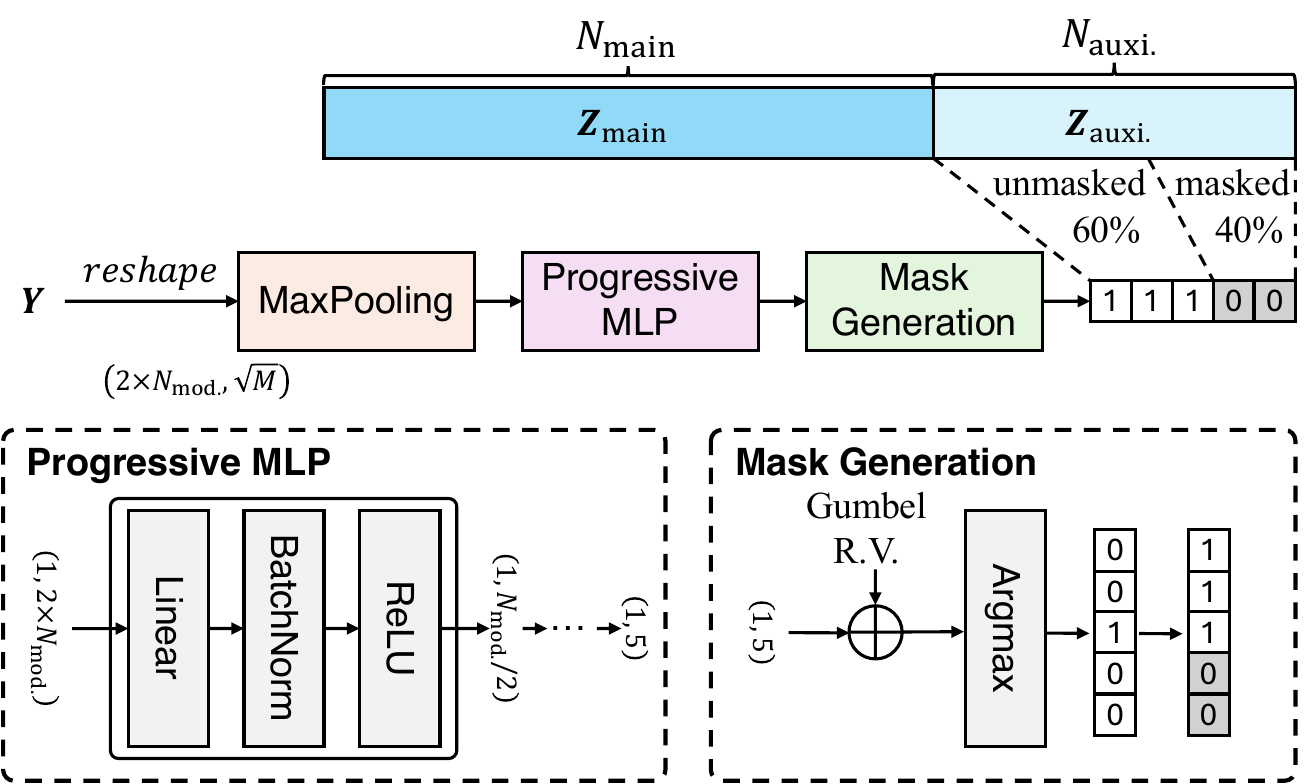}
        \caption{Rate Allocator}
        \label{fig: rate}
    \end{subfigure}

    \begin{subfigure}{0.40\textwidth}
        \centering
        \includegraphics[width=\textwidth]{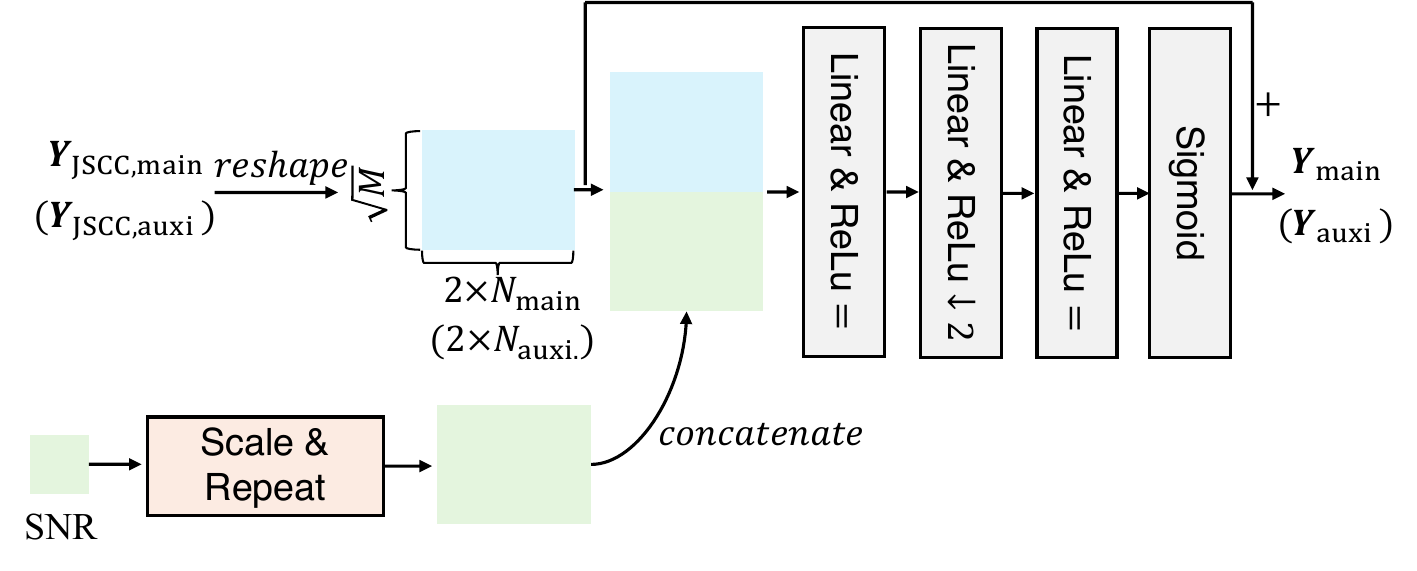}
        \caption{Channel Adapter}
        \label{fig: channel}
    \end{subfigure}
    
    \caption{The model architectures of the rate allocator and channel adapter.}
    \label{fig: adaptive}
\end{figure}

For the rate allocator, motivated by \cite{yang2022deep}, we employ Gumbel-Softmax to achieve differentiable rate selection. Nevertheless, unlike the approach in \cite{yang2022deep}, we do not incorporate channel features into mask generation. Instead, the rate selection is guided solely by $\boldsymbol{Y}$. This enables a decoupling of functional modules. The specific structure of the rate allocator is shown in Fig. \ref{fig: rate}. It first reshapes $\boldsymbol{Y}$ so that the max pooling layer operates along the dimension representing constellation point position probabilities. The pooled result is then passed through a Progressive MLP. Each block reduces the output feature dimension to $\frac{1}{4}$ of its input until the target number of rate levels is reached. Here we set the rate level to $5$. Subsequently, the cutoff position for the mask is determined using the Gumbel-Max method, where the non-differentiable $\operatorname{argmax}$ will be replaced by $\operatorname{softmax}$ during backpropagation, resulting in a one-hot vector. This one-hot vector is further transformed into a thermal vector representing the mask, where $1$ indicates that the corresponding constellation point should be transmitted, and $0$ indicates that it should not. It is worth noting that the rate allocator only adjusts the number of constellation points originating from the auxiliary branch. Because we allocate more modulation symbols and apply an additional Point Transformer Block to the main branch. To ensure high reconstruction quality, we retained all symbols generated by the main branch, which are more informative. Since this masking strategy only discards constellation points at the tail, zero padding can be applied at the receiver without the need to transmit the mask.

For the channel adapter, we fuse the semantic information of the point cloud with channel conditions using the concatenation method. Since the logits output by the designed parallel JSCC encoders represent the probability of constellation point positions, we can directly adjust the logits to influence constellation point generation with a single module at the encoder. This approach contrasts with implicit SNR refinement methods, which typically insert SNR-based feature adjustment modules after each feature extraction layer in both the encoder and decoder \cite{zhang2024analog, yang2024digital}. The detailed structure of the channel adapter is demonstrated in Fig. \ref{fig: channel}. Scale means the SNR will be divided by $10$ to avoid numerical issues.

\subsection{Token Decoder and Point De-tokenizer} 

The token decoder and point de-tokenizer are responsible for converting the modulated tokens received through the channel back into a point cloud, thereby completing the reconstruction task. We primarily adopt the receiver-side model proposed in \cite{bian2024wireless} to implement the token decoder and point de-tokenizer.

Concerning the token decoder, when the transmitter is equipped with a rate allocator, the token decoder first pads the received modulated tokens with zeros to get $\hat{\boldsymbol{Z}}_{\text{pad.}} \in \mathbb{R}^{N_{\text{mod.}} \times 2}$ with a fixed length, enabling further processing by the demodulator. The demodulation process is divided into in-phase and quadrature branches, where two separate $1$D transposed convolution layers are used to adjust both the channel dimension and feature dimension. The semantic features from the two branches are finally merged to obtain $\hat{\boldsymbol{Y}} \in \mathbb{R}^{N_{\text{demod.}} \times \frac{N}{16}}$. In the JSCC decoder, $\hat{\boldsymbol{Y}}$ is transformed into downsampled point coordinates. Based on the downsampled point coordinates, $\hat{\boldsymbol{Y}}$ can be refined by a Point Transformer, resulting in $\hat{\boldsymbol{Y}}_{\text{refined}} \in \mathbb{R}^{\frac{N}{16} \times N_{\text{demod.}}}$. $\hat{\boldsymbol{Y}}_{\text{refined}}$ can estimate point coordinates again, obtaining $\hat{\boldsymbol{X}}_{\text{pt}} \in \mathbb{R}^{\frac{N}{16} \times 3}$.

As for the point de-tokenizer, we utilize the upsampling module in \cite{wiesmann2021deep}. This method upsamples the point cloud by adding different offsets to each point. For instance, for the output of the JSCC decoder $(\hat{\boldsymbol{X}}_{\text{pt}},\hat{\boldsymbol{Y}}_{\text{refined}})$, the upsampled point $\left \{ (\hat{\boldsymbol{X}}^{k}_{\text{pt}},\hat{\boldsymbol{Y}}^{k}_{\text{refined}}) \right \}, k \in \left \{ 0,\cdots,K-1 \right \}$ is generated by
\begin{align}
&\hat{\boldsymbol{X}}^{k}_{\text{pt}} = \hat{\boldsymbol{X}}_{\text{pt}} + r \cdot \Delta_{k}(\hat{\boldsymbol{Y}}_{\text{refined}}), \\
& \hat{\boldsymbol{Y}}^{k}_{\text{refined}} = \Phi_{k}(\hat{\boldsymbol{Y}}_{\text{refined}}),
\end{align}
where $\Delta_{k}, \Phi_{k}, k=0,\cdots,K-1$ are MLPs. The last layer of $\Delta_{k}$ is $\operatorname{tanh}$. $r$ is used to control the offset range. In this point de-tokenizer, two upsampling operations are performed, each with a $\times4$ upsampling ratio, namely $K=4$.

\section{Simulation Results}
\label{sec: Simulation Results}

\subsection{Simulation Settings}
\subsubsection{Dataset}
The publicly accessible dataset ShapeNetCore.v2 \cite{chang2015shapenet}, which contains 55 common objects with about 51000 pre-aligned 3D models, is utilized in the simulations. ShapeNetCore.v2 is split into training, validation, and test sets with a ratio of $70\%$, $10\%$, and $20\%$, respectively. In simulations, the FPS algorithm is used to uniformly sample 2,048 points from the surface of each shape in the dataset to construct the point clouds, namely $N=2048$.

\subsubsection{Training Details}
As for loss functions, when the rate allocator is not included, the Chamfer distance (CD) metric is utilized, representing reconstruction performance. CD is defined as 
\begin{align}
\mathcal{L}_\text{CD}= \frac{1}{|\boldsymbol{X}|} \sum_{\boldsymbol{x} \in \boldsymbol{X}} \min_{\hat{\boldsymbol{x}} \in \hat{\boldsymbol{X}}} \| \boldsymbol{x} - \hat{\boldsymbol{x}} \|_2^2 + \frac{1}{|\hat{\boldsymbol{X}}|} \sum_{\hat{\boldsymbol{x}} \in \hat{\boldsymbol{X}}} \min_{\boldsymbol{x} \in \boldsymbol{X}} \| \hat{\boldsymbol{x}} - \boldsymbol{x} \|_2^2,
\end{align}
which is symmetric, differentiable, and computationally efficient. If the rate adaptive capacity is considered, another term $\mathcal{L}_\text{Rate}=N_\text{mod}/N_\text{send}$ for controlling the transmitted constellation points is added to $\mathcal{L}_\text{CD}$. The loss function can be calculated as
\begin{align}
\mathcal{L} = \mathcal{L}_\text{CD} + \lambda \mathcal{L}_\text{Rate},
\end{align}
where $\lambda$ is a hyperparameter for balancing the reconstruction quality and the number of modulated tokens. We set $\lambda=2\times10^{-4}$ in simulations. 

The batch size in training is $256$. The Adam optimizer is utilized with an initial learning rate $1 \times 10^{-3}$ and weight decay factor $1 \times 10^{-4}$. The learning rate scheduler adjusts the learning rate every $20$ epochs, reducing it to half of its previous value each time. We train our models with the channel adaptive ability within a specific SNR range $[-0.5\text{dB},10.5\text{dB}]$. During training, the SNR value for each batch in an iteration is uniformly sampled from $[-0.5\text{dB},10.5\text{dB}]$. Signal power normalization is performed with respect to all batches within the iteration. $P_{\text{signal}}$ is normalized to $1$. As for the Rayleigh fading channel, the channel gain $h$ is created as $h\sim \mathcal{CN}(0,1)$. Across all experiments, the models under Rayleigh fading channels are derived by fine-tuning the models trained under AWGN channels with the same SNR settings.

\subsubsection{Benchmarks}
To demonstrate the effectiveness of our proposed methods, we choose baseline methods introduced as follows.
\begin{itemize}
  \item \textit{``Modulated SEPT'' method}: SEPT \cite{bian2024wireless} is a DL-based analog JSCC method for point clouds. Since this method produces analog outputs, to ensure a fair comparison, we apply the differentiable modulation method proposed in Section \ref{JSCCM} to SEPT \cite{bian2024wireless}. 

  \item \textit{``Modulated PCST'' method}: We refer to the point semantic communication method in \cite{xie2024semantic} as PCST (Point Cloud Semantic Transmission). \cite{xie2024semantic} considered a higher bitrate range and calculated the bit rate by assuming that each floating-point number is represented with 16 bits. For a fair comparison, we employed the differentiable modulation method proposed in Section \ref{JSCCM} to perform modulation and control the bit rate to the same range. We reproduced PCST following \cite{xie2024semantic}: the number of points per patch was set to 256, the dimension of global information was 4, and the dimension of local semantic information was increased from 8 to 16. The MVTorch \cite{Hamdi2024MVTN} library was used to perform projection, with the number of projection maps set to 4. The number of symbols for the lossless transmission part was computed using the Shannon channel capacity formula.

  \item \textit{``Soft Quantization'' method}:  Soft Quantization is a digital JSCC technology proposed in \cite{tung2022deepjscc}. It generates channel symbols based on the softmax weighted sum of distances from JSCC outputs to predefined constellation sets in backpropagation. To ensure that the employed Soft Quantization in this baseline is fully consistent with ours, we omitted the temperature coefficient adjustment during training in \cite{tung2022deepjscc}. To control the number of transmitted symbols, our JSCC applies max pooling when using the Soft Quantization, resulting in features of size $N_{\text{mod.}} \times 2$.
  
  \item \textit{``Gumbel-Softmax'' method}: Gumbel-Softmax based differential modulation method is proposed in \cite{bo2024joint}. Gumbel-Max is used to perform reparameterized sampling so that constellation points can be generated. Gumbel-Softmax is adopted during backpropagation to handle the non-differentiability of the $\operatorname{argmax}$ operation. 
  
  \item \textit{``STE'' method}: STE is a simple gradient estimation method that directly sets the output of a non-differentiable operation equal to its input in the backward phase. In \cite{yang2024digital}, STE is applied to differentiable modulation. In this paper, models based on the STE method are finetuned on models trained with the Soft Quantization method.
  
  \item \textit{``Uniform Noise'' method}: This method approximates the non-differentiable quantization through adding uniform noise to the quantization inputs in backpropagation \cite{zhang2024analog}. In this paper, models using the Uniform Noise method are also finetuned on Soft Quantization models.
  
  \item \textit{``G-PCC + LDPC'' method}: G-PCC is a point cloud geometry compression standard proposed by Moving Picture Experts Group (MPEG). We use the command-line tool \textit{mpeg-pcc-tmc13} \footnote{[Online]. Available: \url{https://github.com/MPEGGroup/mpeg-pcc-tmc13}} proposed by MPEG to realize the G-PCC codec. The parameters of encoding and decoding are specified as \cite{wang2021multiscale}, which meet MPEG common test conditions. LDPC is chosen for channel coding. We implement it through Sionna \cite{hoydis2022sionna}, an open-source library for link-level simulations based on TensorFlow. The selection of coding rate and modulation order follows Table 5.1.3.1-1 for the Physical Downlink Shared Channel (PDSCH) in 3GPP TS 38.214 version 16.2.0 \footnote{[Online]. Available: \url{https://www.etsi.org/deliver/etsi_ts/138200_138299/138214/16.02.00_60/ts_138214v160200p.pdf}}. In the simulations, the modulation and coding scheme (MCS) that yields the best performance is selected.

\end{itemize}

\subsubsection{Details and Abbreviation of Our Methods}
All the temperature hyperparameter $T$ used is set to $1.5$. We consider cases where $N_{\text{mod.}} = 50, 100, 150, 200, 250, 300$. 
Except for $ N_{\text{mod.}} = 300 $, where the ratio between $ N_{\text{main}} $ and $ N_{\text{auxi.}} $ is set to $2:1$, the ratio is $4:1$ in all other cases. In all proposed models, $N^{'}=512$ and $N^{''}=128$. For the MLP in the main JSCC encoder, the output dimension of each linear layer is $[128,128,2\sqrt{M}]$. While for the MLP in the auxiliary JSCC encoder, the output dimension is $[128,128,2\sqrt{M}]$. For MLPs $\gamma$ and $\delta$ in Point Transformer layers, output dimensions are the same as input feature dimensions. 

The proposed token communication system with JSCCM in the token encoder is named PointTC (Point Token Communications). The optional rate allocator and channel adapter are abbreviated as RA and CA, respectively.

\subsection{Performance Analysis on Various SNRs and Rates}
\begin{figure*}[!t]
    \centering
    \begin{subfigure}{0.75\textwidth}
        \centering
        \includegraphics[width=\textwidth]{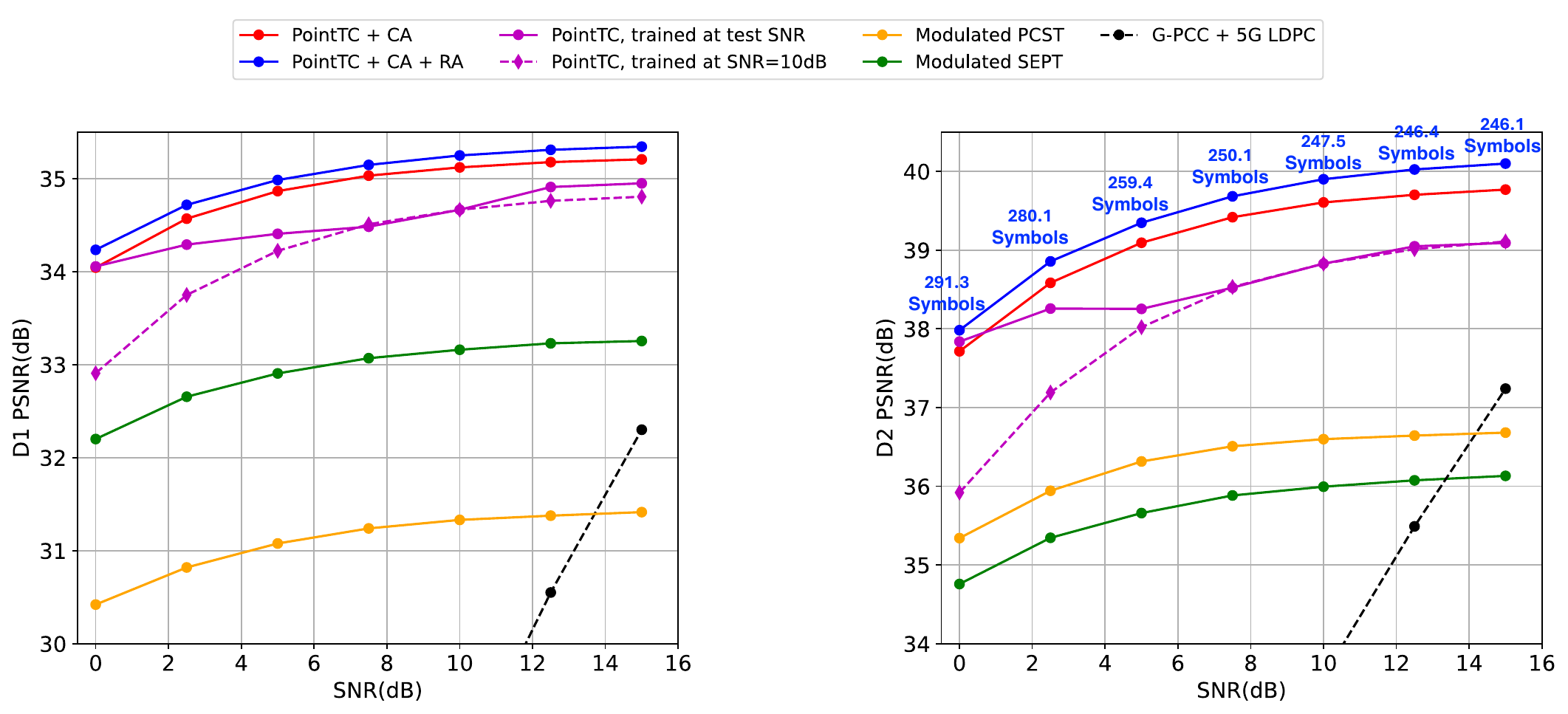}
        \caption{AWGN}
        \label{fig: DvsSNR_AWGN}
    \end{subfigure}

    \begin{subfigure}{0.75\textwidth}
        \centering
        \includegraphics[width=\textwidth]{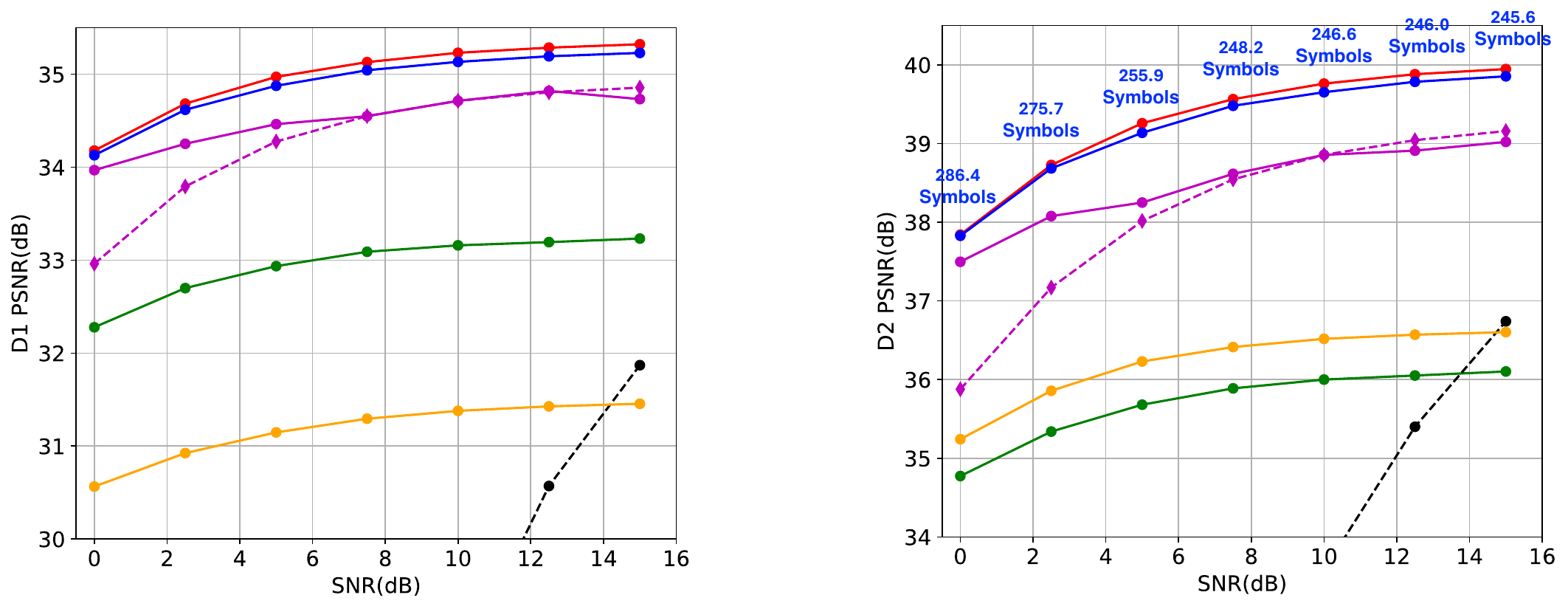}
        \caption{Rayleigh Fading}
        \label{fig: DvsSNR_Rayleigh}
    \end{subfigure}
    
    \caption{The reconstruction performance versus SNRs over AWGN channels and Rayleigh fading channels. The blue numbers on the D2-PSNR subplots indicate the actual number of modulated tokens transmitted by the PointTC+CA+RA model when $N_{\text{mod.}}$ = 300 is configured.}
    \label{fig: DvsSNR}
\end{figure*}

Fig. \ref{fig: DvsSNR} demonstrates the reconstruction performance over AWGN and Rayleigh fading channels under various SNRs. Except for G-PCC+LDPC, all methods in this figure use $300$ channel symbols with 64-QAM, whereas PointTC+CA+RA can adaptively adjust the number of modulated tokens to be transmitted. The simulation results indicate that the proposed methods outperform Modulated SEPT and the traditional separated scheme G-PCC+LDPC under all SNRs. The separated digital methods suffer from a severe cliff effect, while the proposed method still achieves satisfactory performance at 0 dB despite using standard digital constellation points.

Besides, in Fig. \ref{fig: DvsSNR}, the PointTC trained at 10 dB under AWGN channels exhibits inferior reconstruction performance compared to PointTC trained and tested at matched SNRs, with a performance gap exceeding 1 dB when SNR is 0 dB in terms of D1 PSNR. For D2 PSNR, the performance gap between the two methods narrows in the high-SNR regime, as D2 PSNR does not emphasize the precise reconstruction of geometric details. However, a noticeable gap remains in the low-SNR regime. Moreover, PointTC trained at matched SNRs requires multiple separate models, whereas PointTC+CA achieves superior performance using only a single model by adjusting constellation point positions through a concise channel adapter. With CA, the model can learn more robust and informative features under varying training SNRs. Notably, compared to PointTC+CA, PointTC+CA+RA further improves performance under AWGN channels. Because using a fixed transmission rate for a single model under varying channel conditions is suboptimal. PointTC+CA+RA adopts a strategy of transmitting more symbols under poorer channel conditions to achieve better end-to-end reconstruction performance, thereby achieving a better balance between communication efficiency and reconstruction quality across different SNRs. As for Rayleigh fading channels, since perfect channel state information is assumed and equalization is performed, the model refined on AWGN channels exhibits similar performance on Rayleigh fading channels. In Fig. \ref{fig: DvsSNR_Rayleigh}, PointTC+CA+RA underperforms PointTC+CA. This may be because RA operates without explicit channel information. Its ability to infer channel conditions must be learned implicitly through backpropagation, which is more difficult under Rayleigh fading. Moreover, complex channel conditions degrade the effectiveness of RA’s probabilistic sampling, impeding the learning of an appropriate cutoff position.

\begin{figure*}[!t]
    \centering
    \includegraphics[width=0.75\textwidth]{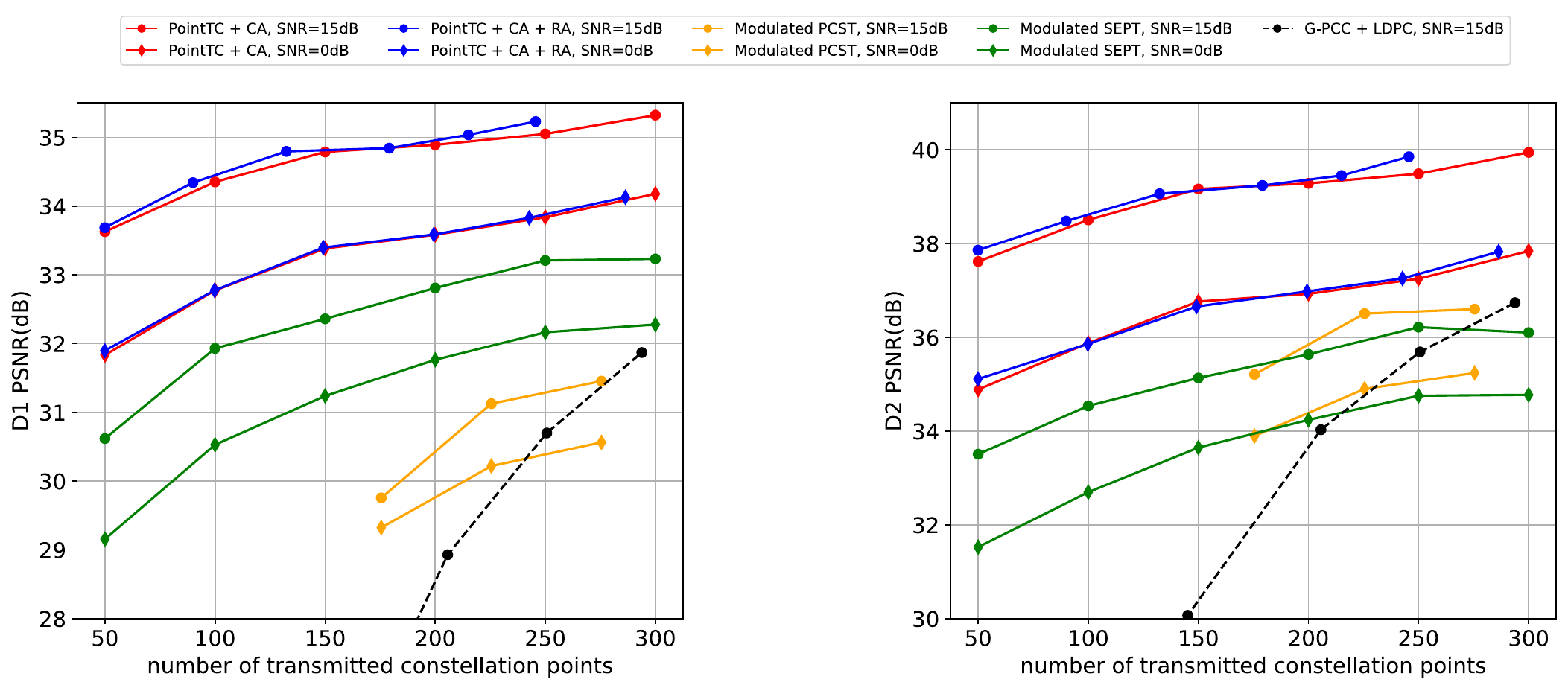}
    \caption{The reconstruction performance versus the number of transmitted constellation points under Rayleigh fading channels.}
    \label{fig: DvsR_Rayleigh}
\end{figure*}

Fig. \ref{fig: DvsR_Rayleigh} depicts D1 PSNR and D2 PSNR performance under Rayleigh fading channels when the number of transmitted symbols varies. We present performance curves under 15 dB and 0 dB. For traditional separation-based methods, bit errors at 0 dB prevent the G-PCC decoder from successfully decoding. In contrast, all methods implemented with JSCCM can still complete decoding at 0 dB. In terms of D1 PSNR, modulated SEPT outperforms the separation-based approach even when the test SNR is 0 dB. However, for D2 PSNR, when the number of transmitted constellation points exceeds 200, the separation-based method begins to surpass modulated SEPT. In comparison, our proposed PointTC-based methods demonstrate consistently superior performance across both D1 PSNR and D2 PSNR metrics, achieving a compression ratio exceeding 6$\times$ in terms of transmitted constellation points.

In Fig. \ref{fig: DvsR_Rayleigh}, the D2 PSNR performance of Modulated PCST is superior to its D1 PSNR performance. This is because PCST adopts a strategy that captures point cloud projections using virtual cameras and then performs feature fusion. The projection features in PCST are assumed to be transmitted losslessly, which preserves the global structural information, leading to better D2 PSNR. However, under the same number of transmitted symbols, PCST cannot maintain fine geometric details, resulting in inferior D1 PSNR performance. PCST is more suitable for cases with a larger number of transmitted symbols. When sufficient transmission symbols are available, the proportion of symbols used for the lossless transmission becomes relatively small, allowing more symbols to be allocated for local features to present geometric details.

\begin{figure}[!t]
    \centering
    \includegraphics[width=0.35\textwidth]{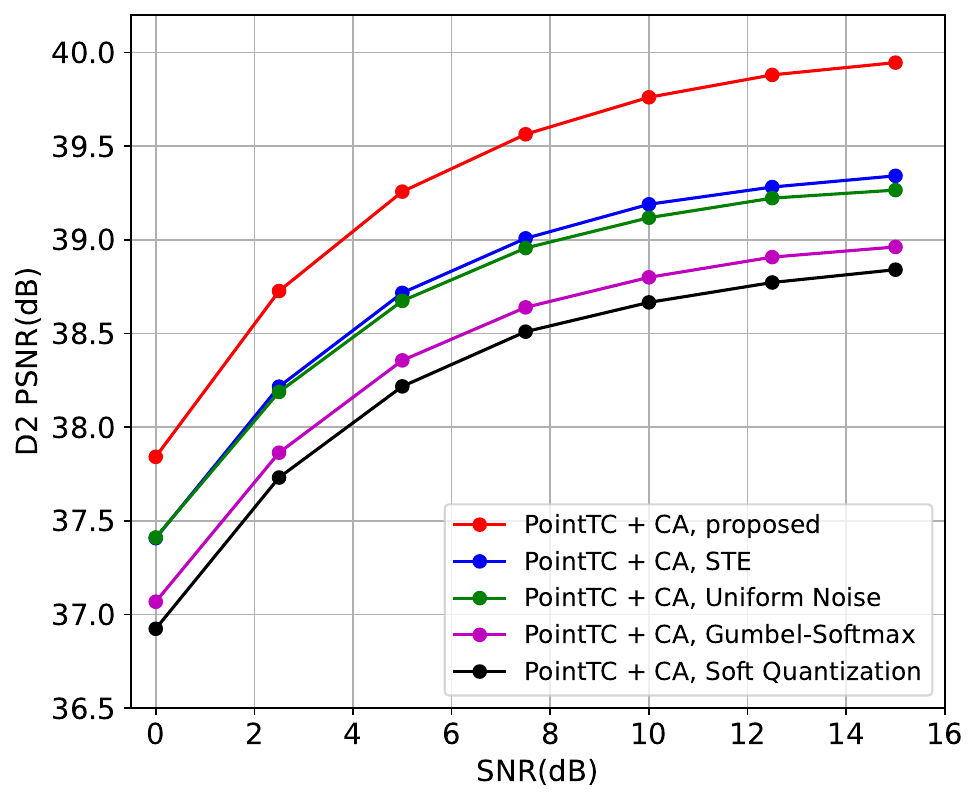}
    \caption{The D2 PSNR performance versus SNRs under Rayleigh fading channels for different modulation methods.}
    \label{fig: modulation}
\end{figure}

\subsection{Performance Analysis on Modulation Methods}
In Fig. \ref{fig: modulation}, we compare the proposed JSCCM scheme with other modulation methods commonly used in digital semantic communication. For fair comparison, we adopt 64-QAM for all methods and fix the number of transmitted constellation points to $300$. It can be observed that our modulation approach, which combines Gumbel-Softmax and soft quantization, consistently outperforms existing differentiable modulation methods across all SNRs. Interestingly, the most intuitive method STE, achieves the best performance among the baselines. Although the Uniform Noise method theoretically provides a better approximation of the quantization process, its performance in our experiments is comparable to that of STE, possibly requiring more sophisticated training strategies for further improvement. When compared to the proposed modulation method, except at 0 dB, either the Gumbel-Softmax-only or the Soft-Quantization-only method shows limited representational capacity, incurring a PSNR drop of more than 1 dB.

\begin{figure}[!t]
    \centering
    \includegraphics[width=0.35\textwidth]{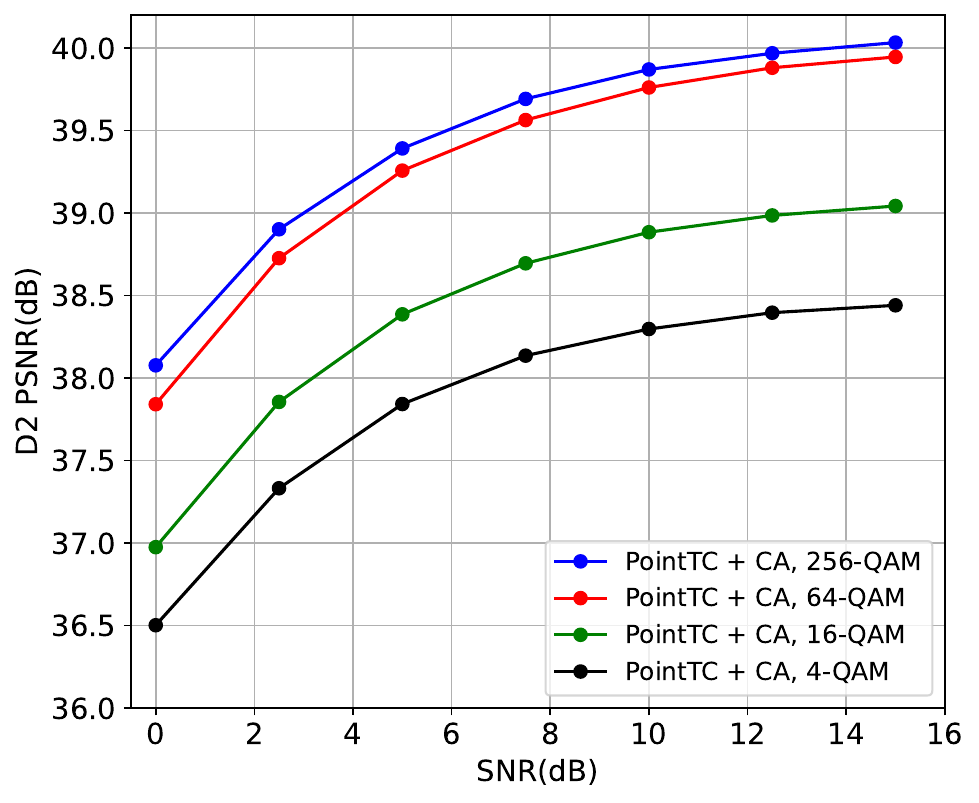}
    \caption{The D2 PSNR performance versus SNRs under Rayleigh fading channels of the proposed JSCCM with different modulation orders.}
    \label{fig: QAM}
\end{figure}
In Fig. \ref{fig: QAM}, we evaluate the performance of the proposed JSCCM under different modulation orders, with the number of transmitted constellation points fixed at $300$. It is observed that, across all SNRs, higher modulation orders lead to better point cloud reconstruction performance. From 4-QAM to 64-QAM, increasing the modulation order results in a performance gain of approximately 1.5 dB. However, the improvement becomes marginal as the modulation order increases beyond 64-QAM, indicating that 64-QAM is approaching the performance limit of the employed JSCC method. This experiment also suggests that, regardless of the SNR condition, selecting the model with the highest modulation order yields the best performance. This contrasts with the conclusion of traditional adaptive modulation, which selects lower modulation orders under low SNR conditions. This difference arises from the fact that traditional adaptive modulation aims to optimize digital communication systems with respect to metrics such as bit error rate. Using high-order modulation under low SNR typically leads to demodulation errors. In contrast, our proposed method treats the received constellation points as input features to a neural network at the receiver side, without converting them into bits. As a result, higher modulation orders provide more expressive features for the decoder, leading to better representation capability.

\subsection{Visualization Results}
\begin{figure*}[!t]
    \centering
    \includegraphics[width=0.75\textwidth]{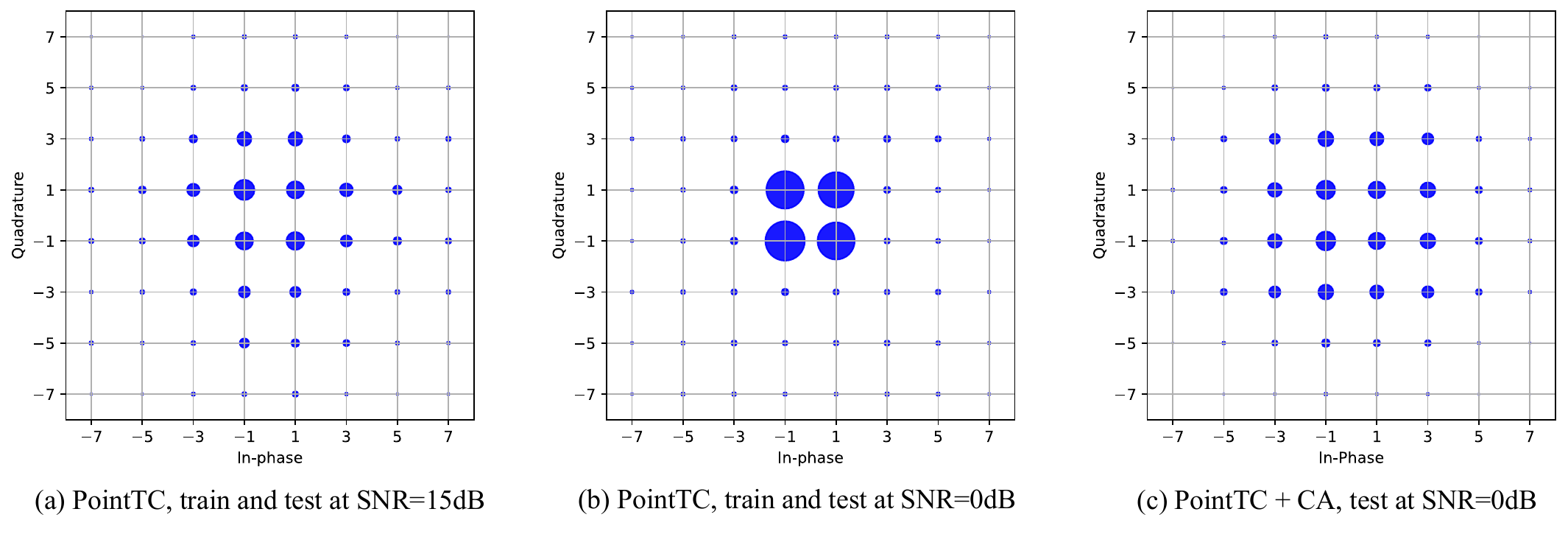}
    \caption{Constellation point distributions generated by different models. The size of the blue circle indicates the probability of constellation points occurring at that location. The larger the circle, the higher the probability.}
    \label{fig: constellation}
\end{figure*}
We analyzed the constellation point distributions produced by the token encoders of different models over 10,000 point clouds from the test set, with the visualization results shown in Fig. \ref{fig: constellation}. It can be observed that the models trained at the test SNR exhibit constellation distributions consistent with the conclusions of traditional adaptive modulation. Namely, lower modulation orders are selected under worse channel conditions. However, the results in Fig. \ref{fig: DvsSNR} have demonstrated that models trained at the test SNR perform worse than the PointTC+CA model. Notably, PointTC+CA does not adopt a lower modulation order at 0 dB. Instead, it maintains a relatively high modulation order, providing the token decoder at the receiver with richer semantic information, which in turn leads to better performance. Furthermore, regardless of the model used, the constellation points produced by the proposed JSCCM method exhibit a distribution resembling a two-dimensional independent Gaussian. Although no prior distribution was explicitly imposed, the model achieves a form of probabilistic shaping gain \cite{forney2003efficient} through end-to-end training.

\begin{figure}[!t]
    \centering
    \includegraphics[width=0.40\textwidth]{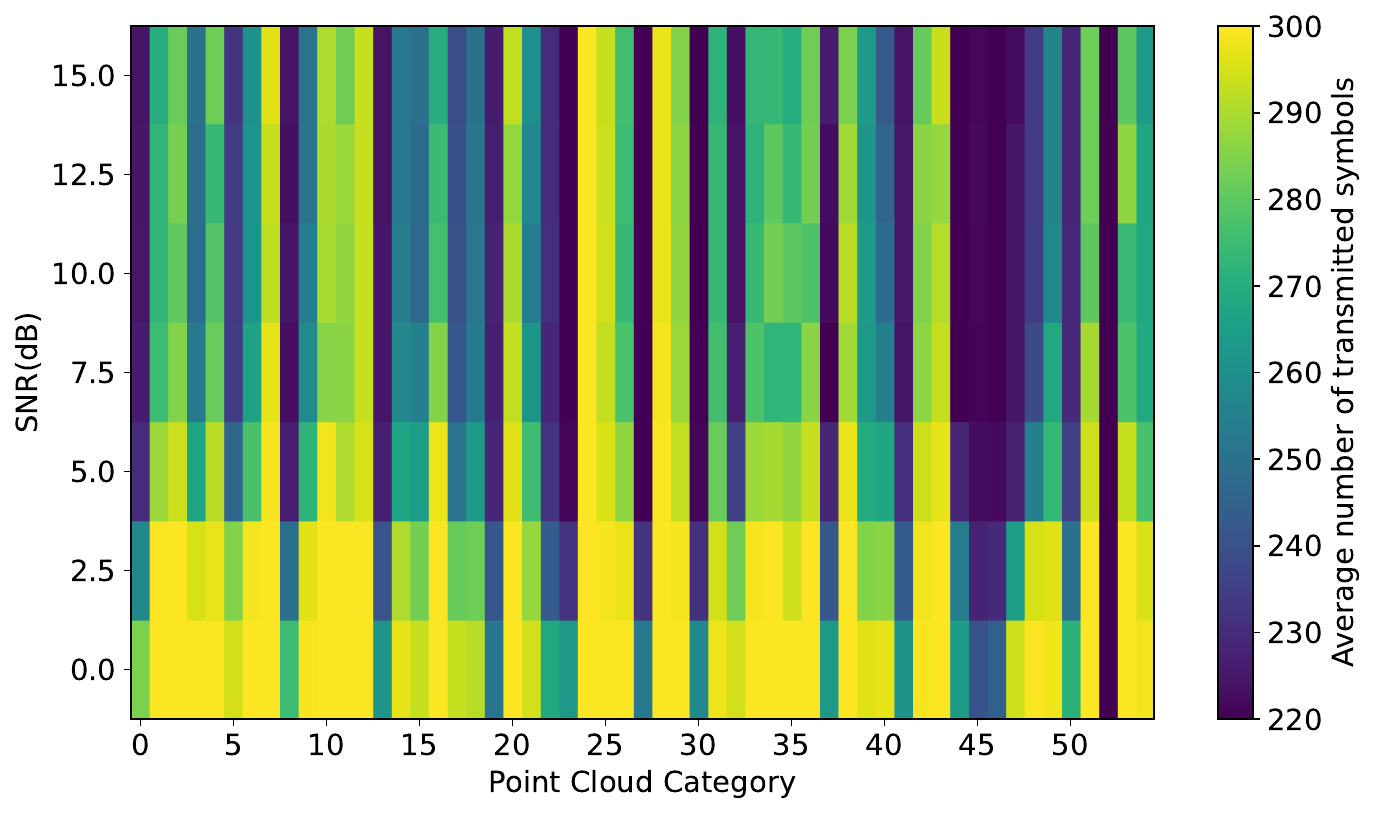}
    \caption{The average number of transmitted symbols versus various SNRs and point cloud categories.}
    \label{fig: heatmap}
\end{figure}

\begin{figure}[!t]
    \centering
    \includegraphics[width=0.40\textwidth]{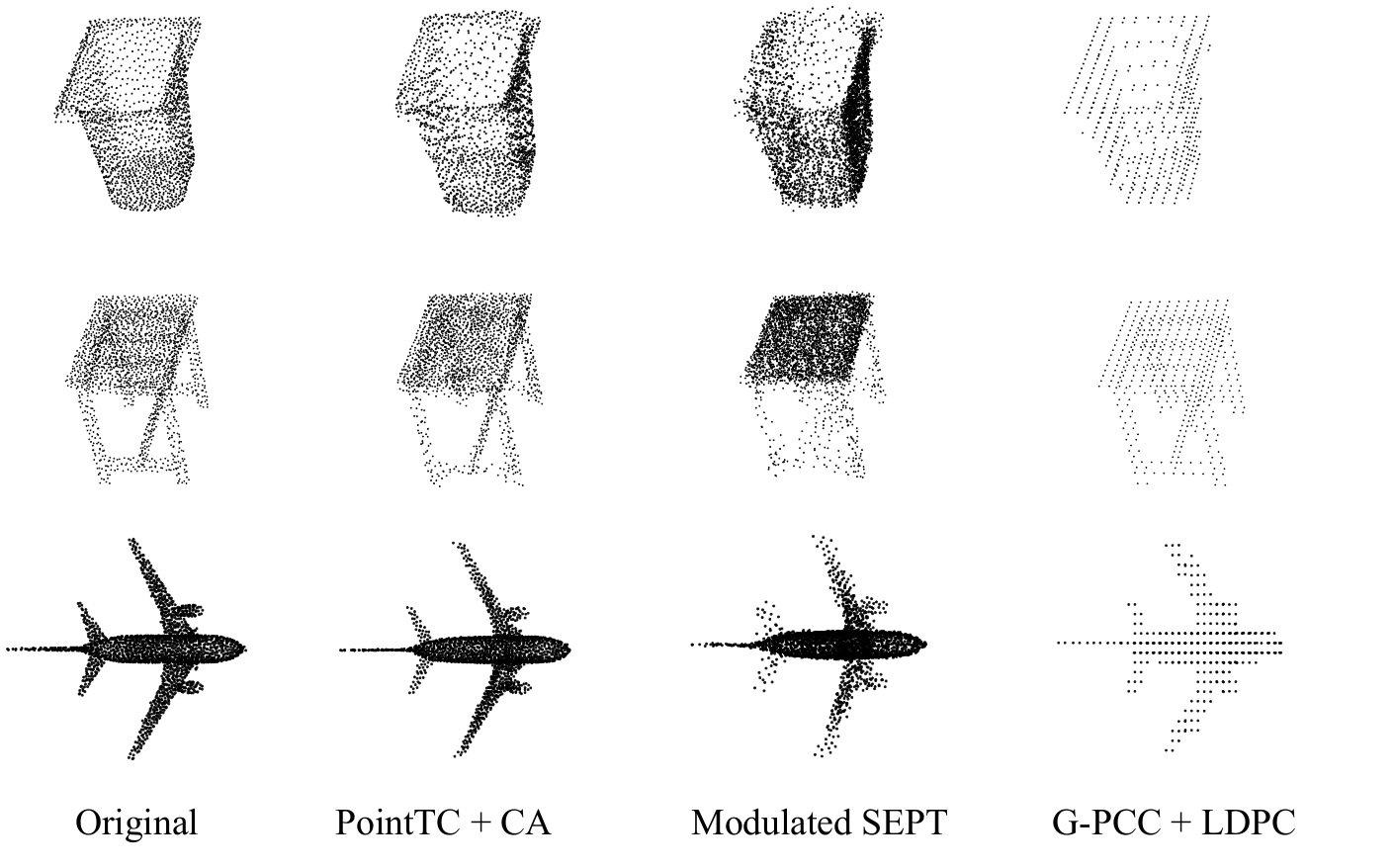}
    \caption{Visualization results of reconstruction.}
    \label{fig: point}
\end{figure}

The number of transmitted symbols for different types of point clouds under various SNRs is shown in Fig. \ref{fig: heatmap}. As illustrated, there is a substantial variation in the number of transmitted symbols across different point cloud categories under the same SNR, indicating that the RA module effectively allocates modulated tokens based on the point cloud semantics. For the same point cloud category, a clear trend emerges that as the SNR increases, the number of transmitted symbols decreases. This demonstrates that both the RA and CA modules contribute to the token communication system’s ability to adapt to both semantic content and channel conditions. Visualization results of the point cloud reconstruction performance are shown in Fig. \ref{fig: point}. We fix the number of transmitted constellation points to 300 and conduct testing at an SNR of 15 dB. It can be observed that the proposed method shows a clear advantage over the baselines in both reconstruction details and subjective visual quality.
\subsection{Performance Analysis on Practical Scenarios} 
\begin{figure}[!t]
    \centering
    \includegraphics[width=0.48\textwidth]{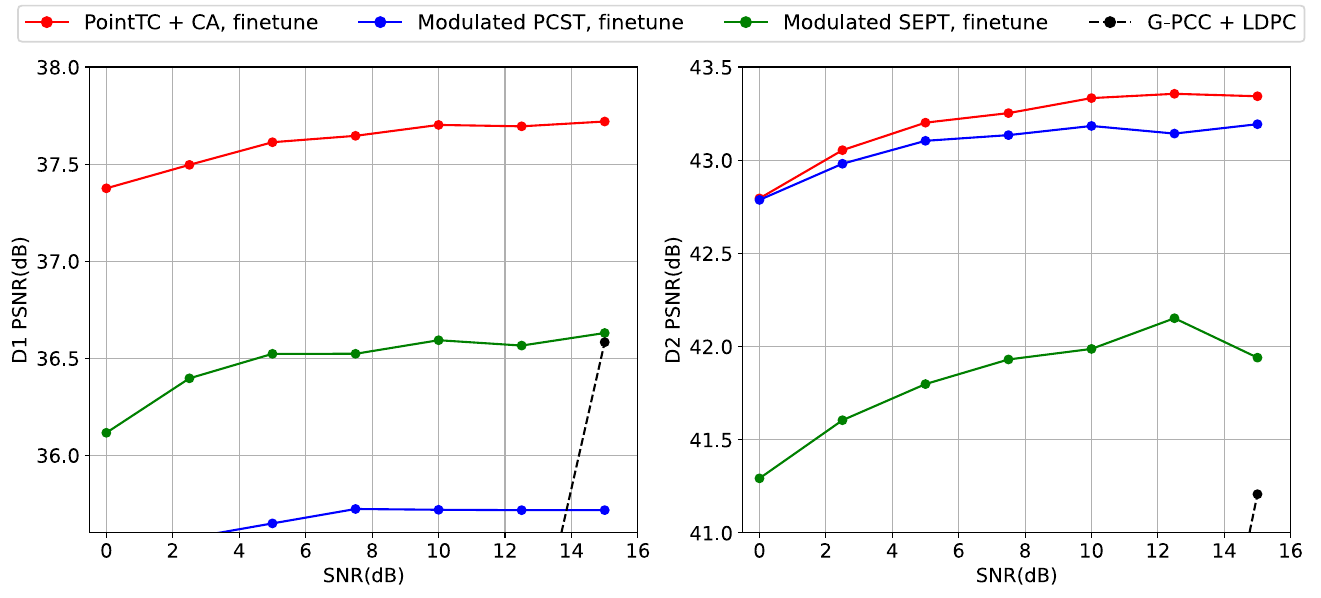}
    \caption{The reconstruction performance versus SNRs over Rayleigh fading channels on SemanticKITTI dataset.}
    \label{fig: KITTI}
\end{figure}

Considering that the point clouds used in the previous simulations were generated by sampling from common 3D models in the ShapeNetCore.v2 dataset, we further evaluated our model on the real-world point cloud dataset SemanticKITTI \cite{Behley2019SemanticKITTI}. We followed the data processing procedure in \cite{wiesmann2021deep}. The performance of the PointTC + CA model trained on ShapeNetCore.v2 is presented in Table \ref{tab1_KITTI}, where a substantial degradation in both D1 PSNR and D2 PSNR can be observed. This degradation occurs because the ShapeNetCore.v2 dataset consists of synthetic point clouds, whereas the SemanticKITTI dataset contains real point clouds captured by vehicle-mounted Light Detection and Ranging (LiDAR) sensors. To address this issue, we further finetuned the PointTC + CA model on the SemanticKITTI dataset to enhance its performance. The corresponding results are also listed in Table \ref{tab1_KITTI}, and the comparison between the finetuned model and the baseline is shown in Fig. \ref{fig: KITTI}. It can be observed that, after finetuning, our model achieves a significant performance improvement and outperforms baselines.

\begin{table}[t!]
\centering
\renewcommand{\arraystretch}{1.2}
\caption{Reconstruction performance on SemanticKITTI. Each entry shows D1 PSNR / D2 PSNR (dB).}
\resizebox{1\columnwidth}{!}{
\begin{tabular}{c|cccc}
\hline
                                                   & \textbf{0.0 dB} & \textbf{5.0 dB} & \textbf{10.0 dB} & \textbf{15.0 dB}     \\ \hline
\textbf{PointTC+CA}                                & 15.2 / 22.7     & 15.2 / 23.0     & 15.1 / 23.2      & 15.0 / 23.2          \\
\multicolumn{1}{l|}{\textbf{PointTC+CA, finetune}} & 37.4 / 42.8     & 37.6 / 43.1     & 37.7 / 43.3      & 37.7 / 43.3 \\ \hline
\end{tabular}}
\label{tab1_KITTI}
\end{table}

\begin{figure}[!t]
    \centering
    \includegraphics[width=0.35\textwidth]{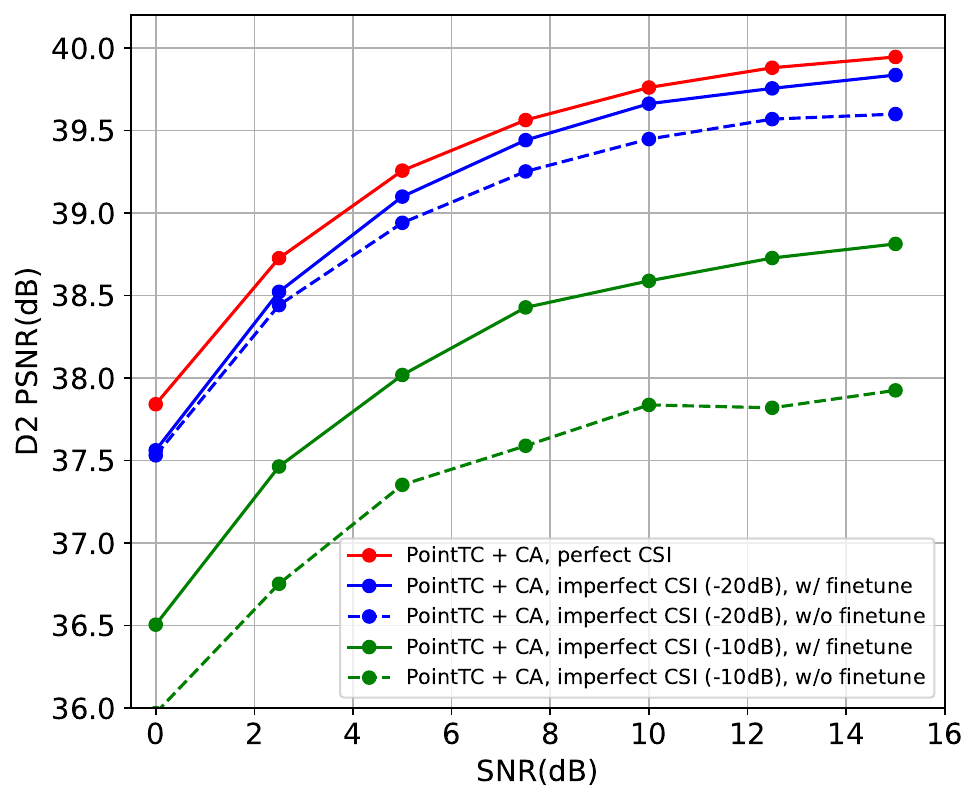}
    \caption{The D2 PSNR performance versus SNRs under perfect CSI and imperfect CSI conditions.}
    \label{fig: CSI}
\end{figure}

In the previous simulations, for the Rayleigh fading channel, we assumed that the ZF equalizer had perfect channel state information (CSI), which is difficult to achieve in practical applications. To evaluate the model performance under imperfect CSI conditions, we added two levels of complex Gaussian white noise $n\sim \mathcal{CN}(0,\sigma_n^2)$ to the channel gain $h$. $\sigma_n^2=10^{-2}$ (-20 dB) represents a good CSI estimation while $\sigma_n^2=10^{-1}$ (-10 dB) represents a poor CSI estimation. The results are shown in Fig. \ref{fig: CSI}. It can be observed that, with -20 dB noise, the model experiences only about a 0.3 dB performance loss, which is acceptable. However, with -10 dB noise, the performance degradation exceeds 1.5 dB. To improve performance under imperfect CSI, we finetuned the model under CSI with noise. As shown in Fig. \ref{fig: CSI}, the finetuned model exhibits reduced performance loss under imperfect CSI, and the improvement becomes more pronounced under better channel conditions. These results demonstrate that our model possesses strong robustness against imperfect CSI estimation results.

\begin{table}[t!]
\centering
\renewcommand{\arraystretch}{1.2}
\caption{Computational and Model Complexity.}
\resizebox{1\columnwidth}{!}{
\begin{tabular}{c|cccc}
\hline
\textbf{Models}            & \textbf{Enc time (ms)} & \textbf{Ded time (ms)} & \textbf{FLOPs (G)} & \textbf{Parameters (M)} \\ \hline
\textbf{PointTC + CA}      & \textbf{12.8}          & 1.2                    & \textbf{10.06}     & 24.45                   \\
\textbf{PointTC + CA + RA} & 15.0                   & 1.3                    & \textbf{10.06}     & 24.47                   \\ \hline
Modulated SEPT             & 55.4                   & 1.4                    & 15.38              & 12.33                   \\
Modulated PCST             & 19.5                   & \textbf{0.2}           & 11.06              & \textbf{5.96}           \\
GPCC + LDPC                & 34.2                   & 568                    & -                  & -                       \\ \hline
\end{tabular}}
\label{tab2_complexity}
\end{table}

Finally, we conducted experiments to evaluate the computational and model complexity. The results obtained on an NVIDIA RTX 4090 GPU are presented in Table \ref{tab2_complexity}. The results show that our model achieves the lowest end-to-end codec latency and requires fewer floating-point operations than the baselines. With an end-to-end latency of 14 ms, the PointTC + CA model can achieve a 71.4 frame per second (FPS) codec speed. This efficiency is attributed to the fact that, although the Point Transformer architecture introduces a relatively large number of parameters, GPUs are well optimized for Transformer-based structures. In addition, the time-consuming kNN and FPS operations are implemented using CUDA in our models, which further reduces the overall codec time. These results demonstrate the strong potential of our models for practical applications.

\section{Conclusion and Future Work}
\label{sec: Conclusion}
In this paper, we have developed a token wireless communication system for point cloud transmission. To obtain informative and robust token representations, we design a joint semantic-channel coding and modulation scheme for the token encoder, which consists of two parallel Point Transformer-based JSCC encoders and a differential modulator. The differential modulator combines Gumbel-softmax and soft quantization methods to generate high-quality modulated tokens. Additionally, the rate allocator and channel adapter are based on the physical meaning of JSCC outputs. Simulation results demonstrate that our proposed methods exhibit superior performance compared to the existing DL-enabled JSCC method and traditional separated codec.

In essence, this paper presents a promising framework for token communication in point clouds. However, many open problems remain to be explored.
\begin{itemize}
\item In this paper, we only consider point geometry and static point clouds. Further studies are needed on point attributes and dynamic point clouds transmission for more immersive applications.
\item Our models require finetuning for real points. Developing methods tailored for real-world data with varying point densities and acquisition noise is necessary.
\item Although this paper proposes the JSCCM framework for token communication, models are designed for point clouds. Building a unified token communication system for multimodal data is an important direction.
\end{itemize}

\bibliographystyle{IEEEtran}
\bibliography{reference}

\end{document}